\documentclass[
 aps,
 prb,
 reprint,
 amsmath,
 amssymb,
 superscriptaddress
 ]{revtex4-2}

\usepackage{cochineal}
\usepackage{color}
\usepackage{array}
\usepackage{units}
\usepackage{mathrsfs}

\usepackage{graphicx}

\usepackage{hyperref}
 \hypersetup{
     colorlinks   = true,
     citecolor    = blue,
     linkcolor    = blue,
     urlcolor     = blue
}

\makeatletter

\PassOptionsToPackage{version=3}{mhchem}
\usepackage{mhchem}

\newcommand{\TSDW}{T_\text{SDW}}
\newcommand{\TN}{T_\text{N}}
\newcommand{\Tc}{T_\text{c}}

\newcommand{\Euxii}{\ce{Eu_{0.73}Ca_{0.27}(Fe_{0.87}Co_{0.13})2As2}}
\newcommand{\Euviii}{\ce{Eu(Fe_{0.81}Co_{0.19})2As2}}
\newcommand{\Eu}{\ce{EuFe2As2}}

\newcommand{\cax}{\ensuremath{c}\nobreakdash-axis}
\newcommand{\abplane}{\ensuremath{ab}\nobreakdash-plane}

\newcommand{\MS}{M\"ossbauer spectroscopy}
\newcommand{\Moss}{M\"ossbauer}

\newcommand{\Feion}{\ce{Fe^{2+}}}
\newcommand{\Euion}{\ce{Eu^{2+}}}

\newcommand{\UH}{U_\text{H}}
\newcommand{\RH}{R_\text{H}}
\newcommand{\Ro}{R_\text{o}}
\newcommand{\Rs}{R_\text{s}}

\newcommand{\Hcr}{H_\text{cr}}
\newcommand{\Hac}{H_\text{ac}}

\newcommand{\muO}{\mu_0}
\newcommand{\Thp}{\theta_\text{p}}

\newcommand{\dd}{\textrm{d}}

\newcommand{\oC}{{}^\circ C}

\graphicspath{Rys/}

\makeatother

\begin{document}

\title{Re-entrance of resistivity due to the interplay \\
of superconductivity and magnetism\\
in \ce{Eu_{0.73}Ca_{0.27}(Fe_{0.87}Co_{0.13})2As2}}
\author{Lan Maria Tran}
\email[Corresponding author: ]{l.m.tran@intibs.pl}

\affiliation{Institute of Low Temperature and Structure Research, Polish Academy of Sciences}
\author{Andrzej J. Zaleski}
\affiliation{Institute of Low Temperature and Structure Research, Polish Academy of Sciences}
\author{\framebox{Zbigniew Bukowski}}
\affiliation{Institute of Low Temperature and Structure Research, Polish Academy of Sciences}
\date{\today}
\begin{abstract}
    By simultaneous Co and Ca-doping we were able to obtain an $\Eu$-based compound with superconductivity appearing above the antiferromagnetic order of $\Euion$ magnetic moments. However, as soon as the antiferromagnetic order appears a re-entrance behavior is observed \textemdash{} instead of zero resistivity and diamagnetic signal down to the temperature of \unit[2]{K}. By investigating magnetization, ac susceptibility and electrical transport properties of $\Euxii$ and comparing them to previously studied \Moss{} effect and neutron scattering measurements of this and similar compounds an explanation of such behavior is proposed.
\end{abstract}

\maketitle

\section{Introduction}
Understanding of the relationship between magnetism and superconductivity in high-temperature superconductors is one of the remaining problems of the solid state physics. While superconductivity and magnetism are considered antagonistic phenomena, on the contrary there are several groups of materials exhibiting both properties simultaneously \citep{SCinTernaryI,SCinTernaryII,Kolodziejczyk1980,Kolodziejczyk1985,Kadowaki1987,Maple1995,Lynn1997,Muller2001b,Ruthenate-book,Nachtrab2006,OrganicSC-book,Pfleiderer2009,Niewa2011,Aoki2011,Knebel2011,Bochenek2015}. Among them the iron based superconductors are an extraordinary group, where the magnetic order and superconductivity originate from the same (iron) sublattice.

It was shown, that by suppression of the spin density wave (SDW) order (associated with the magnetic order on the Fe-sublattice) superconductivity in $M\ce{Fe2As2}$ (where $M=$ $\ce{Ca}$, $\ce{Ba}$, $\ce{Sr}$, $\ce{Eu}$) compounds can be achieved. This can be a result of partial chemical substitution at either the $M$ \citep{Rotter2008PRL,Wu2008,Yuan2009,He2010a}, $\ce{Fe}$ \citep{Sefat2008,Jiang2009PRB,He2010a,Ying2010PRB,Matusiak2011PRB,Blachowski2011Codoped,Harnagea2011,Nicklas2011SCES,KrugvonNidda2012,Anand2015} or $\ce{As}$ \citep{Ren2009PRL,Jeevan2011PRB} sites.

Due to the presence of the localized magnetism of $\Euion$ ions, we find the $\Eu$-based compounds as one of the most interesting among the iron-based pnictide superconductors. In our previous study we focused on the properties of either Co or Ca doped $\Eu$ compounds \citep{Tran2012NJP,Tran2012PRB,Tran2018}. Based on these and other literature results (see Ref. \onlinecite[and references therein]{Zapf2017}) it can be concluded that by diluting Fe-by-Co the temperature of the spin density wave ($\TSDW$) order associated with the Fe-sublattice can be decreased and simultaneously superconductivity induced. Such substitution did not change the temperature of antiferromagnetic ordering of magnetic moments of $\Euion$ ions ($\TN$), however as shown by \Moss{} and neutron spectroscopies, the type of magnetic order on the Eu-sublattice changes significantly with decreased Co\nobreakdash-concentration \citep{Blachowski2011Codoped,Jin2016}. On the other hand, Eu\nobreakdash-by\nobreakdash-Ca substitution decreased $\TN$,  however such systems are not superconducting under ambient pressure \citep{Mitsuda2010JPSJL-p,Mitsuda2011JPSJ,Mitsuda2011SCES,Harnagea2018,Shrestha2020} and while the magnetic order on Eu-sublattice does not change compared to the $\Eu$ parent compound, the magnetic order on Fe\nobreakdash-sublattice is modified below $\TN$ \citep{Tran2018,Jin2019}.

Nevertheless, we believed that in the case of already superconducting systems, superconductivity can be enhanced if the antiferromagnetic order associated with Eu can be weakened or destroyed. Thus we expected that by additional Ca-doping we can obtain a superconducting system with lower $\TN$ and higher $\Tc$ compared to investigated by us $\Euviii$ \citep{Tran2012NJP,Tran2012PRB} and possibly a system where superconductivity appears above Eu-magnetic order. Therefore, this study focuses on the properties of a double-doped compound \textendash{} $\Euxii$ \textendash{} with both Fe and Eu-sublattices diluted by a non-magnetic ions.

\section{Experimental}

The single crystals of $\Euxii$ were grown using the Sn-flux method. High purity elements Eu:Ca:Co:Fe:As:Sn in nominal ratios 0.6:0.4:0.5:1.6:2:30 were inserted in an alumina crucible placed in a quartz tube, than evacuated and flame sealed. The ampule was heated with approximately $\unit[100]{\unitfrac{\oC}{h}}$ heating ratio up to $\unit[1100]{\oC}$ and held at this temperature for about $\unit[10]{h}$, to enable proper dissolving of the components. The crystals grew while the solution was cooled with $1$-$\unit[2]{\unitfrac{\oC}{h}}$ ratio down to $\unit[600]{\oC}$. At this stage the liquid tin was decanted from the crucible. The remanent Sn was dissolved using diluted hydrochloric acid.

The chemical composition of the grown single crystals was determined using EDS spectroscopy. The crystal structure and phase purity of the samples was characterized by powder X-ray diffraction (XRD) using X'Pert Pro powder diffractometer equipped with a linear PIXcel detector and CuK$\alpha$ radiation.

To determine the properties of $\Euxii$ we measured electric transport (resistivity, magnetoresistivity and Hall effect), magnetization and ac susceptibility in $2$-$\unit[300]{K}$ temperature range and in external magnetic fields up to $\unit[9]{T}$. Measurements were performed for fields applied parallel and perpendicular to the crystallographic \cax{} (except Hall effect measurements where fields were only applied parallel to the \cax).

Electric transport and magnetization were investigated using Quantum Design's Physical Properties Measurement System (PPMS) platform. Resistivity (and magnetoresistivity) was measured using the 4-point technique. Silver electrodes were attached to the sample's surface with DuPont silver paste. The measurements were carried out in fields both parallel and perpendicular to the \cax, while the ac electric current $I=\unit[10]{mA}$ with a frequency of $\unit[47]{Hz}$ was always applied parallel to the \abplane.

For Hall effect measurements additional two contacts were attached on the \abplane. To determine the Hall voltage $\UH$ taking into account the error due to the misalignment of the Hall-electrodes, the voltage on the Hall-electrodes was measured while the sample was in magnetic field parallel and antiparallel to the \cax. The Hall voltage was than calculated as $\UH=(U_{\text{H},180^{\circ}}-U_{\text{H},0^{\circ}})/2$, where $U_{\text{H},180^{\circ}}$ and $U_{\text{H},0^{\circ}}$ are the voltage values measured while the sample was rotated by $180^{\circ}$ and $0^{\circ}$, respectively.

The ac susceptibility was measured by applying a driving field of $\muO\Hac=\unit[10]{mT}$ with a frequency of $\unit[1]{kHz}$ using the Oxford Susceptometer or the PPMS with the ACMS option.

In this contribution we present results obtained for one single crystal (unless stated otherwise), however all measurements were also performed for different crystals obtained from the same batch. Therefore, our conclusions are also based on the results obtained from these additional measurements. Chosen additional results are discussed in more detail in Supplementary Materials \citep{Tran-SuppMat}.

\section{Results}

\subsection{Composition and crystal structure}

\begin{figure}[h]
    \includegraphics[width=0.99\columnwidth]{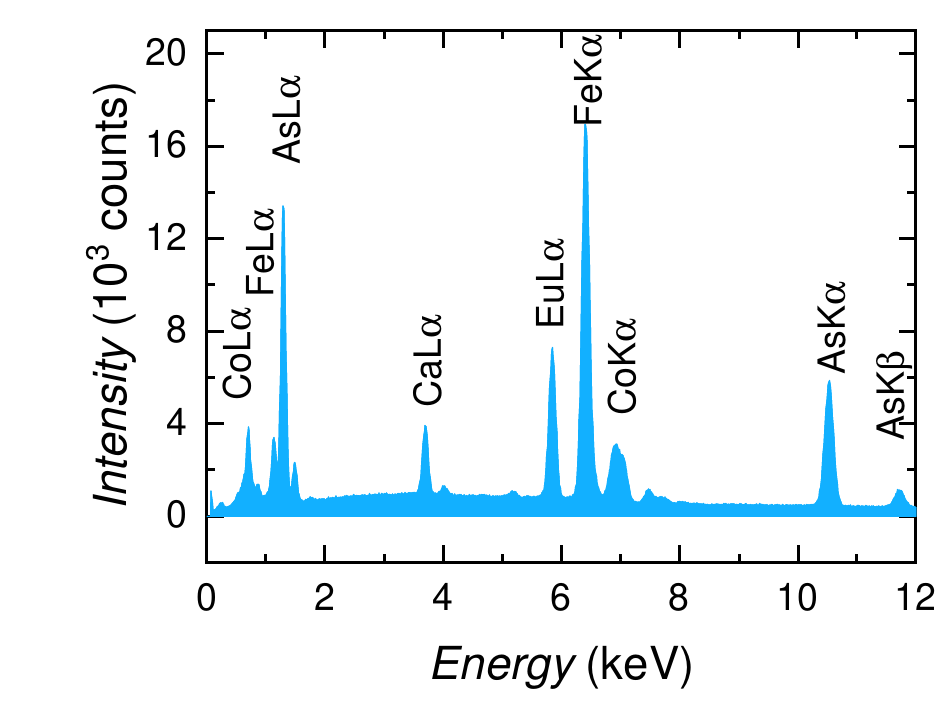}
    \caption{EDX spectrum of \Euxii{}\label{fig:EDX}}
\end{figure}

EDX spectrum of one of the single crystal of \Euxii{} is presented in Fig.~\ref{fig:EDX}. The composition of \Euxii{} was determined taking the average from measurements performed for several samples from the same batch. No traces of tin were detected.

\begin{figure}[h]
    \includegraphics[width=0.99\columnwidth]{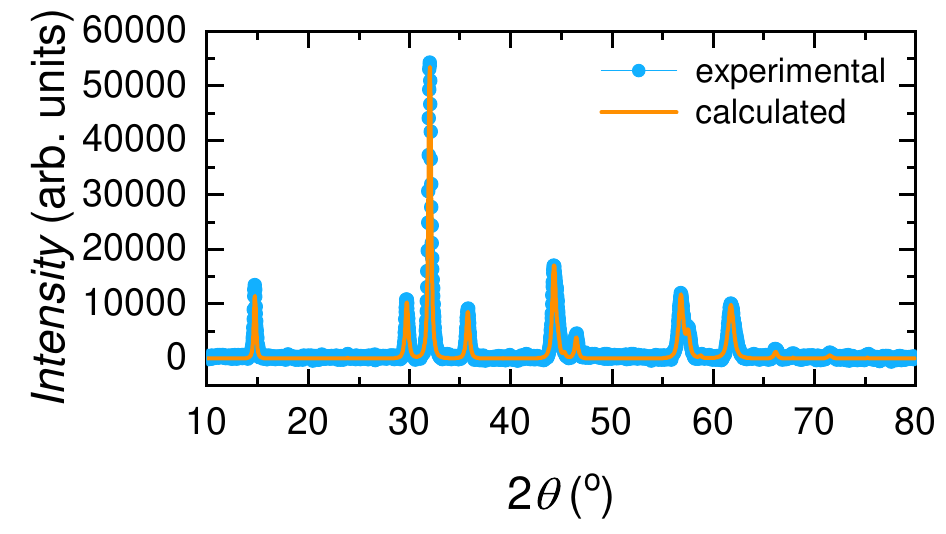}
    \caption{Experimental (blue open circles) and calculated (solid orange lines) room-temperature XRD pattern of powdered crystals of \Euxii{}\label{fig:XRD}}
\end{figure}

Several crystals of \Euxii{} from the same batch were grinded to perform room temperature x-ray powder diffraction measurements. The collected diffraction pattern is presented in Fig.~\ref{fig:XRD}. All the observed reflections for the investigated material could be indexed to the tetragonal \ce{ThCr2Si2}-type structure ($I4/mmm$ space group) expected for the \ce{AFe2As2}-based systems at room temperature. The refined lattice parameters are $a = \unit[3.9075]{\mathring{A}}$ and $c = \unit[12.0092]{\mathring{A}}$. All reflexes could be indexed to the $I4/mmm$ space group, no foreign phases were detected.

\subsection{Electric transport}

\subsubsection{Resistivity and magnetoresistivity}

\begin{figure}[h]
    \includegraphics[width=0.99\columnwidth]{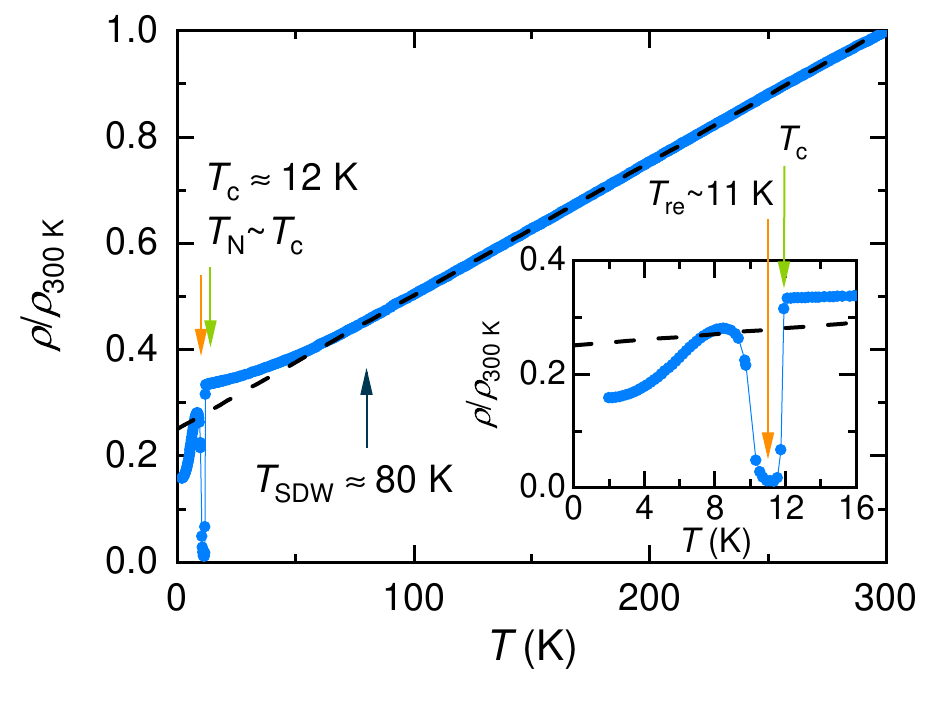}
    \caption{Temperature dependence of resistivity normalized to the resistivity at $\unit[300]{K}$ of $\Euxii$, the inset presents normalized resistivity in $2$-$\unit[16]{K}$ temperature range. The dashed line is the linear extrapolation of the data at temperatures above $\TSDW$ \label{fig:Res(T)}}
\end{figure}

Temperature dependence of (zero-magnetic-field) resistivity normalized to resistivity at $\unit[300]{K}$ $\rho/\rho_{\unit[300]{K}}(T)$ of $\Euxii$ is presented in~Fig.~\ref{fig:Res(T)}, closeup of the $2$-$\unit[16]{K}$ temperature range is shown in the inset.

In the high temperature region resistivity has a linear temperature dependence, down to about $\unit[80]{K}$ where the slope changes (see Fig.~\ref{fig:Res(T)}). High temperature anomalies of the resistivity dependence of the 122-systems are usually associated with the SDW ordering of the Fe $3d$ electrons \citep{Blachowski2011Codoped,Matusiak2011PRB} and as was shown by the \MS{} study such is the case in the investigated compound \citep{Komedera2018}.

However, the most interesting behavior is observed below $\sim\unit[12]{K}$, where the resistivity first rapidly decreases (due to the superconducting transition) and just before zero resistivity is reached, a \textit{re-entrance} of resistivity is observed, i.e., the resistivity increases with decreasing temperature and a resistive state is re-established.

\begin{figure}[h] \hspace{-10pt}
    \includegraphics[width=0.57\columnwidth]{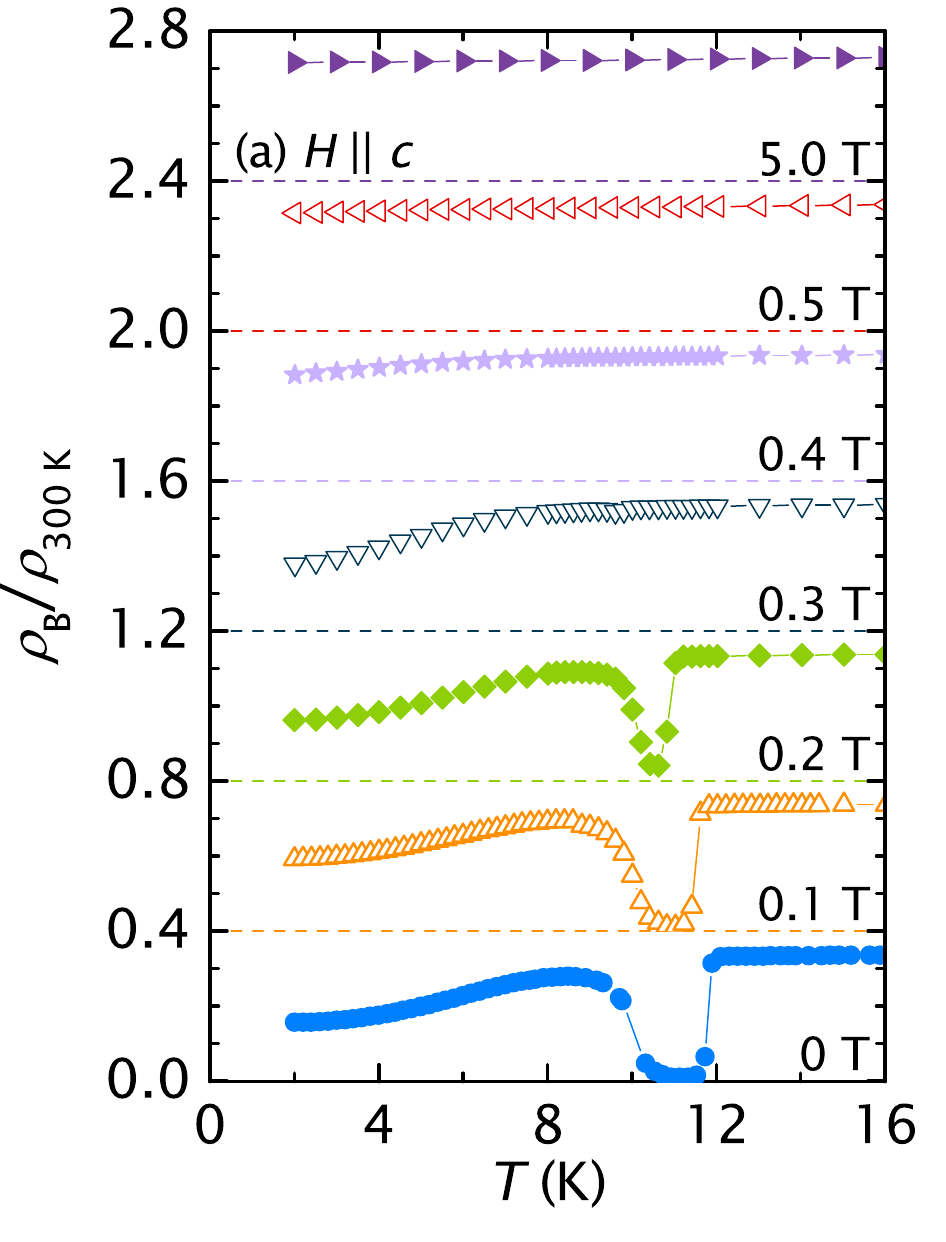}\hspace{-30pt}
    \includegraphics[width=0.57\columnwidth]{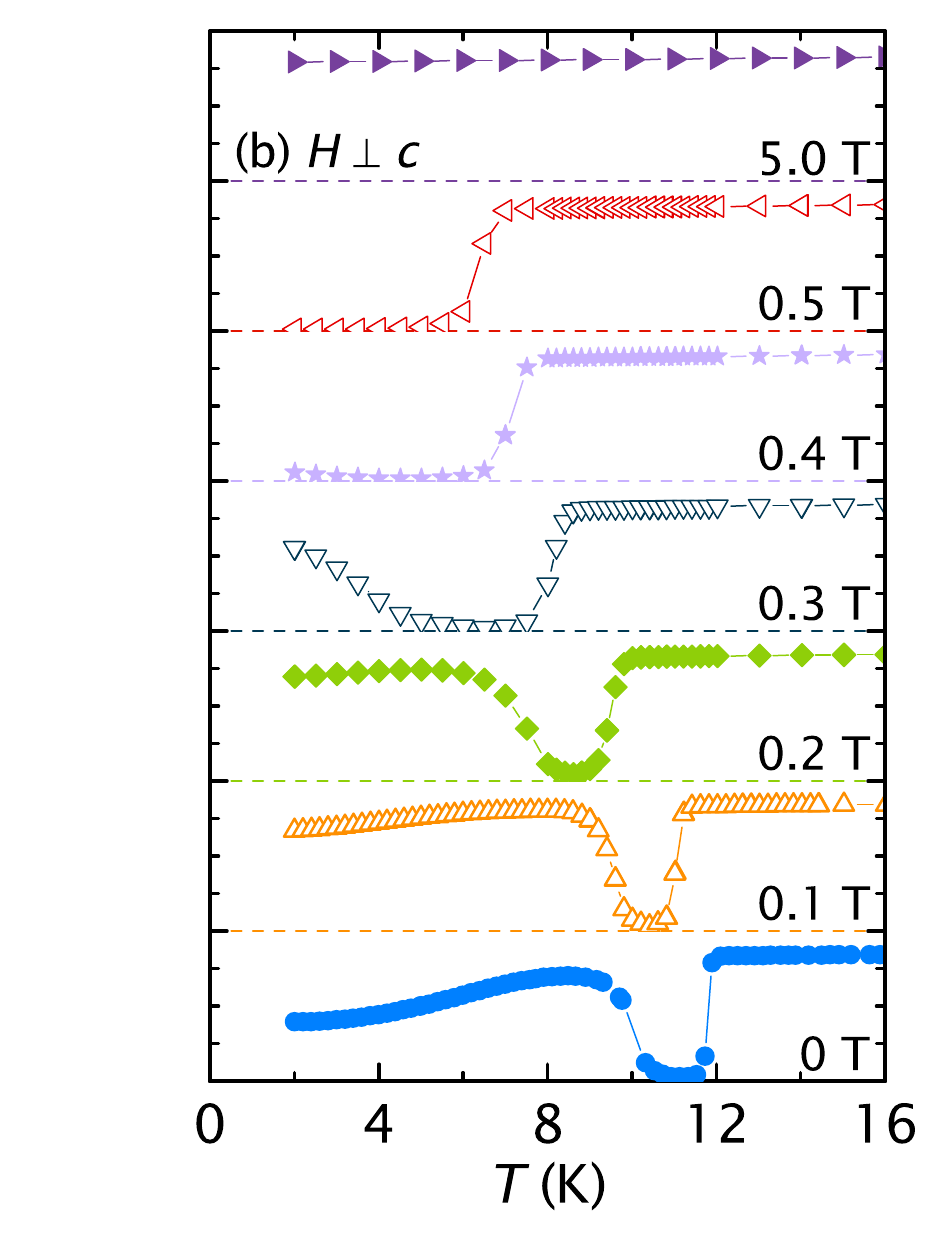}
    \caption{Temperature dependence of magnetoresistivity normalized to $\rho_{\unit[300]{K}}$ measured in external magnetic fields applied (a) parallel and (b) perpendicular to the $c$-axis, note that the dependencies are artificially shifted along the $y$-axis by 0.1 with the $\rho=0$ level marked with dashed lines \label{fig:MR(T)}}
\end{figure}

Measurements carried out in external magnetic fields (see Fig.~\ref{fig:MR(T)}) reveal that the drop of resistivity as well as the re-entrance are shifted to lower temperatures with increasing magnetic fields. However, depending on the direction of the applied field, the temperature dependencies are different. For fields applied parallel to the \cax, the drop of resistivity is smaller in each field and finally can not be distinguished above fields of $\unit[0.3]{T}$. On the other hand, for fields perpendicular to the \cax, the drop can be clearly observed and the superconducting transition is observed up to $\unit[3.5]{T}$, while the ``re-growth'' is not visible at fields above $\unit[0.4]{T}$ in the investigated temperature range. Our explanation of this behavior is presented in Sec.~\ref{sec:Discussion}.

Moreover, it should be noted that resistivity values below the ``re-growth'' are smaller than values expected from extrapolation of normal resistivity above $\unit[12]{K}$ (cf. dashed line in inset of Fig.~\ref{fig:Res(T)}).

From resistivity (magnetoresistivity) measurements onset of the superconducting transition was determined as the temperature at which the resistivity starts to decrease rapidly, the data points are presented on Fig.~\ref{fig:Ph-diag} with solid blue stars (\textsf{\textit{$\textit{\textsf{T}}_{\textit{\textsf{c}}}$}}\textsf{ - SC onset in }\textsf{\textit{$\rho_{\textsf{B}}(\textsf{\textit{\textsf{T}}})$}}). In the same figure with blue open stars (\textsf{\textit{$\textit{\textsf{T}}_{\textsf{reentrance}}$ }}\textsf{in }\textsf{\textit{$\rho_{\textsf{B}}(\textsf{\textit{\textsf{T}}})$}}) are presented data points corresponding to the temperatures at which the resistivity ``re-grows''.

\subsubsection{Hall effect}

\begin{figure}[h]
    \includegraphics[width=0.99\columnwidth]{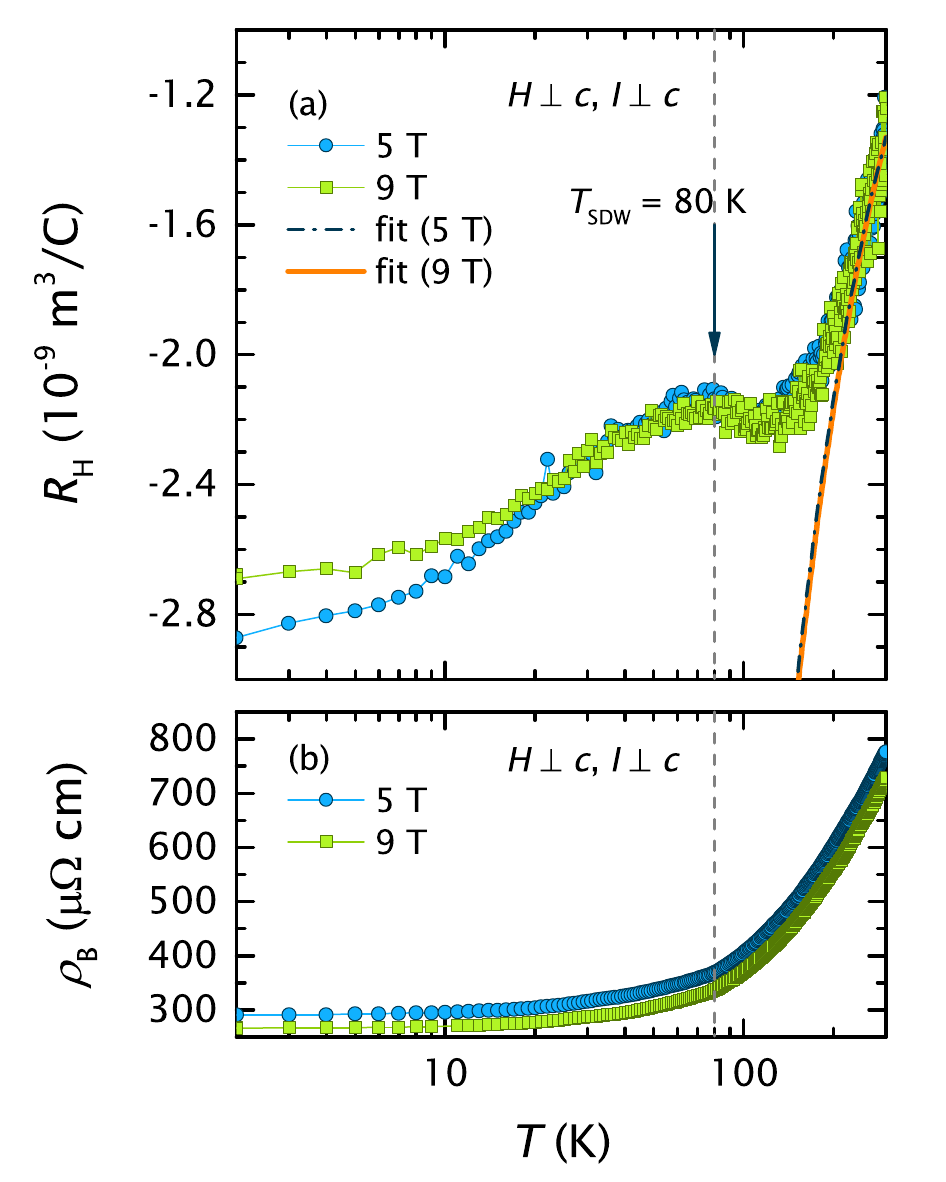}
    \caption{Temperature dependence of (a) Hall coefficient $\RH$ in $5$ and $\unit[9]{T}$ with the fit of Eq.~\ref{eq:RH} and (b) magnetoresistivity at $\unit[9]{T}$. The dashed line represents the SDW transition temperature\label{fig:RH(T)}}
\end{figure}

Hall effect measurements were carried out in fields of $5$ and $\unit[9]{T}$, the respective temperature dependencies are presented in Fig.~\ref{fig:RH(T)}(a). Similarly as was observed for other $\Eu$-based Co-doped compounds,~\citep{Matusiak2011PRB} the Hall coefficient changes with temperature; i.e. at high temperatures $\RH$ decreases with decreasing temperature. While only a slight change of slope was detected in $\rho(T)$ (and $\rho_{B}(T)$) at $\TSDW$ --- see Fig.~\ref{fig:Res(T)} and \ref{fig:RH(T)}(b) \textendash{} a clear anomaly of the $\RH(T)$ is observed at this temperature --- cf. Fig.~\ref{fig:RH(T)}(a). Similar anomaly near $\TSDW$ was found in other $\Eu$-based Co-doped compounds. \citep{Matusiak2011PRB,Tran2012NJP,Ren2008PRB}

In systems with a spin-orbit interactions between the conducting electrons and localized moments, Hall coefficient consist of two components --- the ordinary $\Ro$ and anomalous $\Rs$ contributions --- and can be written as 
\begin{equation} 
    \RH=\Ro+\Rs\chi_{V}(T),\label{eq:RH} 
\end{equation}
where $\chi_{V}(T)$ is the temperature dependent volume susceptibility (in SI units) calculated from magnetization . The fitting of Eq.~\ref{eq:RH} to the experimental data in 250-$\unit[300]{K}$ temperature range is presented as a solid line in Fig.~\ref{fig:RH(T)}. The best fitting parameters are $\Ro=\unit[3.18\cdot10^{-10}]{m^{3}/C}$ and $\Rs=\unit[-3.30\cdot10^{-7}]{m^{3}/C}$ for the $\unit[5]{T}$-measurements, while $\Ro=\unit[3.74\cdot10^{-10}]{m^{3}/C}$ and $\Rs=\unit[-3.42\cdot10^{-7}]{m^{3}/C}$ for the $\unit[9]{T}$-data. A high ratio $\Rs/\Ro$ of order of 1000 suggests a dominance of the spin-orbit coupling in the high-temperature region.

The Hall coefficient is negative over all investigated temperature region, which suggests electrons as majority carriers in $\Euxii$. While it is known that iron based superconductors are multi-band systems,\citep{Zapf2017} with the provided data results it is not possible to separate the components associated with the  hole and electron carriers and only one-band model can be used. This enables crude estimation of the carrier concentration $n$ and mobility $\mu_{\text{H}}$ of the system. In this model, the carrier concentration equals:
\begin{equation}
    n=\frac{1}{e_{0}\Ro},
\end{equation}
where $e_{0}=\unit[1.6\cdot10^{-19}]{C}$ is the value of elementary electric charge; while carrier mobility equals:
\begin{equation}
    \mu_{\text{H}}=|\Ro|\sigma=\frac{|\Ro|}{\rho},
\end{equation}
where carrier conductivity $\sigma$ is defined as inverse resistivity $\nicefrac{1}{\rho}$. The calculated parameters for $\Euxii$ are summarized in Tab.~\ref{tab:Hall}.

\begin{table}[h] 
    \caption{Ordinary and anomalous Hall coefficients, carrier concentration and mobility determined from fitting of Eq.~\ref{eq:RH} to experimental data and calculated in the one-band model
    \label{tab:Hall}}
    \begin{tabular}{>{\raggedleft}p{0.3\columnwidth}>{\centering}p{0.3\columnwidth}>{\centering}p{0.3\columnwidth}} \hline
        {}                                                       & $\unit[5]{T}$       & $\unit[9]{T}$\tabularnewline \hline \hline
        $\Ro$ in $\unitfrac{m^{3}}{C}$                           & $3.18\cdot10^{-10}$ & $3.74\cdot10^{-10}$\tabularnewline \hline
        $\Rs$ in $\unitfrac{m^{3}}{C}$                           & $-3.30\cdot10^{-7}$ & $-3.42\cdot10^{-7}$\tabularnewline \hline
        $n$ in $\unit{\unitfrac{1}{m^{3}}}$                      & $1.96\cdot10^{28}$  & $1.67\cdot10^{28}$\tabularnewline \hline
        $\mu_{\text{H}}$ in $\unit{\unitfrac{cm^{2}}{V\cdot s}}$ & 0.41                & 0.51\tabularnewline \hline
    \end{tabular}
\end{table}

Similar results were obtained for other Co-doped $\Eu$-based compounds.~\citep{Tran2012NJP,Matusiak2011PRB}

\subsection{Magnetization}

\begin{figure}[h] \hspace{-10pt}
    \includegraphics[width=0.57\columnwidth]{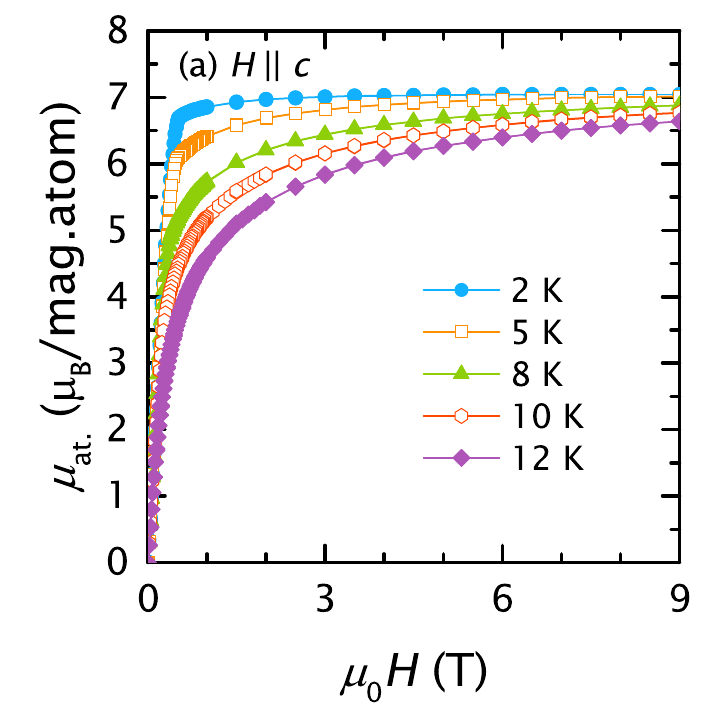}\hspace{-30pt}
    \includegraphics[width=0.57\columnwidth]{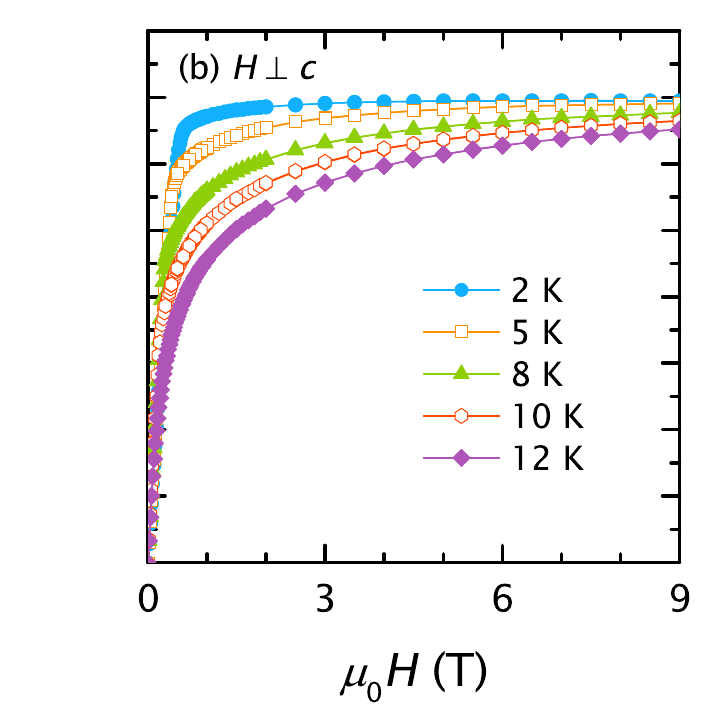}
    \caption{Field dependent magnetization measured at magnetic fields applied (a) parallel and (b) perpendicular to the $c$-axis \label{fig:M(H)}}
\end{figure}

Field dependent magnetization measured in several temperatures is presented in Fig. \ref{fig:M(H)}. No spontaneous magnetization was detected and the initial magnetization increases linearly with increasing external magnetic field, what is expected for an AF-system. All magnetization-curves saturate above a certain magnetic field. The saturation magnetization for the $\unit[2]{K}$\nobreakdash-measurement is around $\mu_{\text{sat}}\approx\unit[7.05]{\mu_{B}}$, which is consistent with the theoretical value for $\Euion$ ion $\mu_{\text{sat}}^{\text{theo}}=g\mathscr{J}\mu_{\text{B}}=\unit[7]{\mu_{B}}$ (where $g=2$ and $\mathscr{J}=\nicefrac{7}{2}$), indicating that all magnetic moments order ferromagnetically (field induced ferromagnetic order).

\begin{figure}[h]
    \includegraphics[width=0.98\columnwidth]{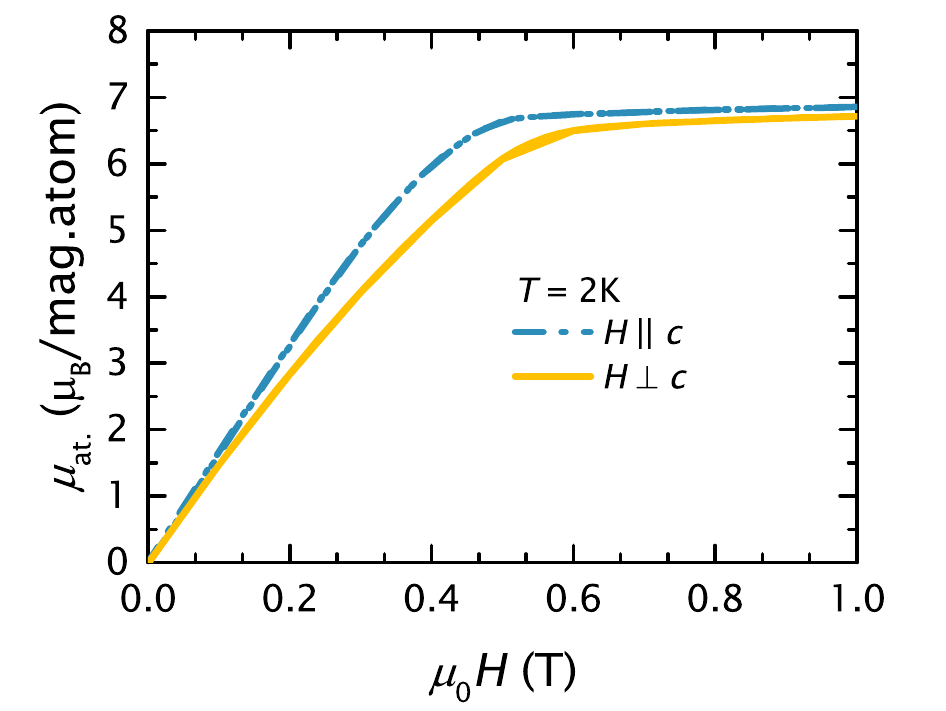}
    \caption{Comparison of the initial slope of field dependent magnetization measured at $\unit[2]{K}$ while the magnetic field was applied parallel and perpendicular to the $c$-axis \label{fig:M(H)-1}}
\end{figure}

Depending on the direction of applied magnetic field, the initial slopes of $M(H)$ are distinctly different with a higher slope value for measurements performed in fields applied parallel to the \cax{} (cf. Fig. \ref{fig:M(H)-1}).  The transition field between AF and field induced ferromagnetic orders $\Hcr$ can be determined by taking the minimum of the second derivative of magnetization $\textrm{d}^{2}M/\textrm{d}H^{2}$, the calculated data points are collected on the magnetic phase diagram in Fig. \ref{fig:Ph-diag} as orange squares.

\begin{figure}[h]
    \includegraphics[width=0.99\columnwidth]{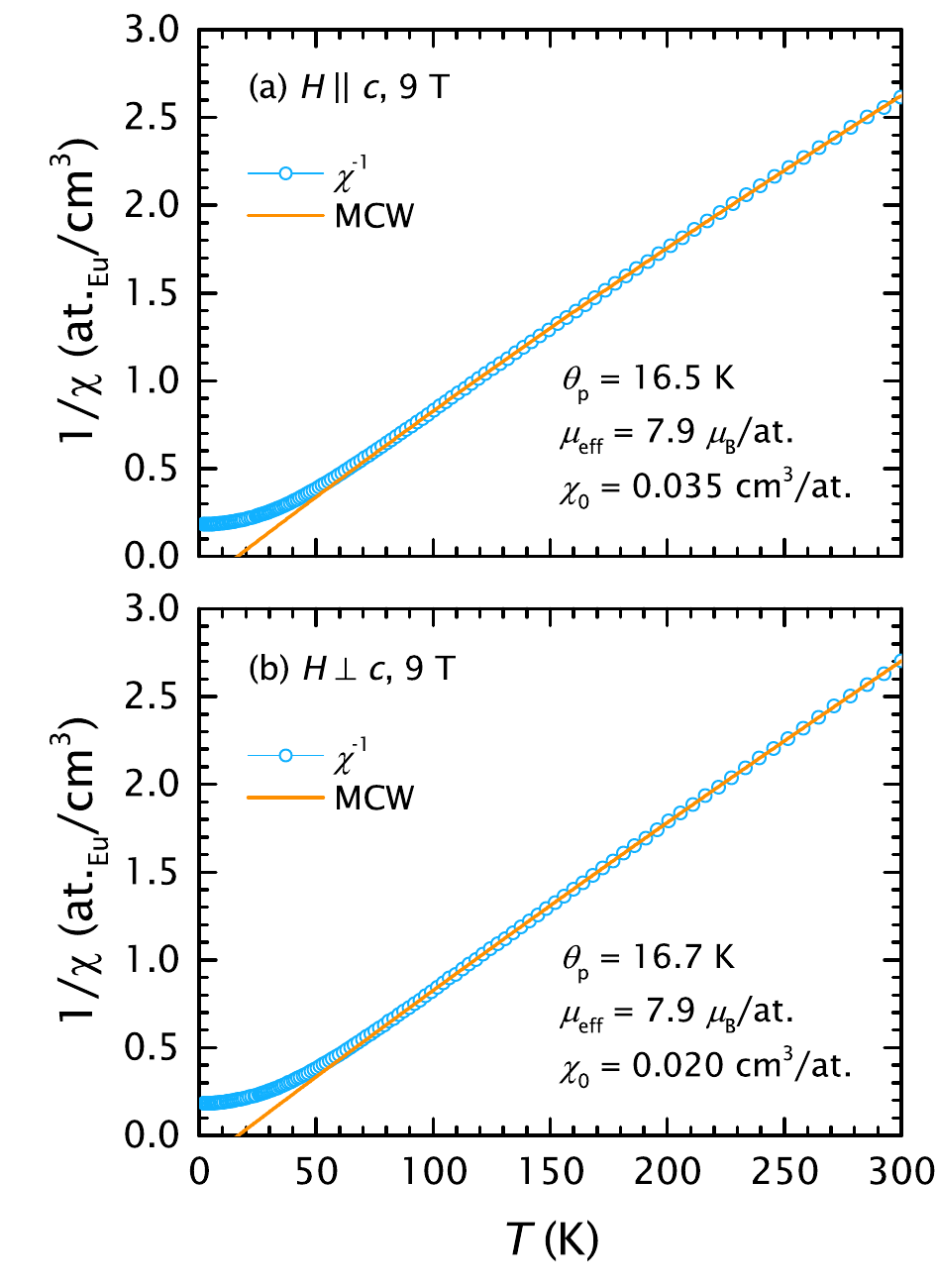}
    \caption{Temperature dependence of inverted dc susceptibility measured in $\unit[9]{T}$ magnetic field applied (a) parallel and (b) perpendicular to the $c$-axis, the solid line represents the fitting of Eq.~\ref{eq:mCW} in $100$-$\unit[300]{K}$ temperature range\label{fig:invChi}}
\end{figure}

Temperature dependence of inverted dc susceptibility investigated at $\unit[9]{T}$ is presented in Fig.~\ref{fig:invChi}. The dependence at high temperatures could be fitted using the modified Curie-Weiss
law:
\begin{equation} 
    \chi(T)=\dfrac{N_{\text{A}}}{3k_{\text{B}}\mu_{0}}\dfrac{\mu_{\text{eff}}^{2}}{T-\Thp}+\chi_{0},\label{eq:mCW}
\end{equation}
where $N_{\text{A}}$ is the the Avogadro's constant, $k_{\text{B}}$ --- the Boltzmann's constant, $\mu_{0}$ --- permeability of free space, $\Thp$ --- the paramagnetic Curie temperature (or Weiss temperature), $\mu_{\text{eff}}$ --- effective magnetic moment (in Bohr magnetons $\mu_{\text{B}}$), and $\chi_{0}$ --- the temperature independent component of susceptibility. The fitting parameters are summarized in Fig.~\ref{fig:invChi}.

The evaluated effective magnetic moments per Eu atom $\mu_{\text{eff}}$ is equal to the theoretical value $\mu_{\text{eff}}^{\text{theo}}=g\sqrt{\mathscr{J}(\mathscr{J}+1)}=\unit[7.9]{\mu_{\text{B}}}$ for divalent $\Euion$ ions. The constant $\chi_{0}$ contribution is negligibly small when compared to the main magnetic susceptibility, confirming the absence of other phases.

While the Weiss temperature $\Thp$ is positive, a ferromagnetic interaction is expected. However, as shown for $\ce{EuCo2As2}$ \citep{Ballinger2012} and multiple $\Eu$-based compounds \citep{Ren2008PRB,Jiang2009NJP,Paramanik2013JPhys,Tran2018,Tran2012NJP,Hu2011} the Weiss temperature is positive although the system is antiferromagnetic. As discussed in Ref.~\citealp{Tran2018} the sign of Weiss temperature is affiliated to the interactions between the nearest-neighbors, therefore the positive value suggest a ferromagnetic interaction among nearest $\Euion$ ions in the \abplane.

The Ne\'{e}l temperature was determined using the Fisher's method \citep{Fisher1962}. For fields applied parallel to the \cax{} the maximum of $\dd\chi(T)T/\dd T$ was not certain, therefore the transition temperature was only estimated for this data set, the procedure how the data points were estimated is described in the Supplementary Materials \citep{Tran-SuppMat}. The evaluated data points are summarized on the magnetic phase diagram in Fig. \ref{fig:Ph-diag} as orange triangles.

\subsection{Susceptibility}

\begin{figure}[h]
    \includegraphics[width=1\columnwidth]{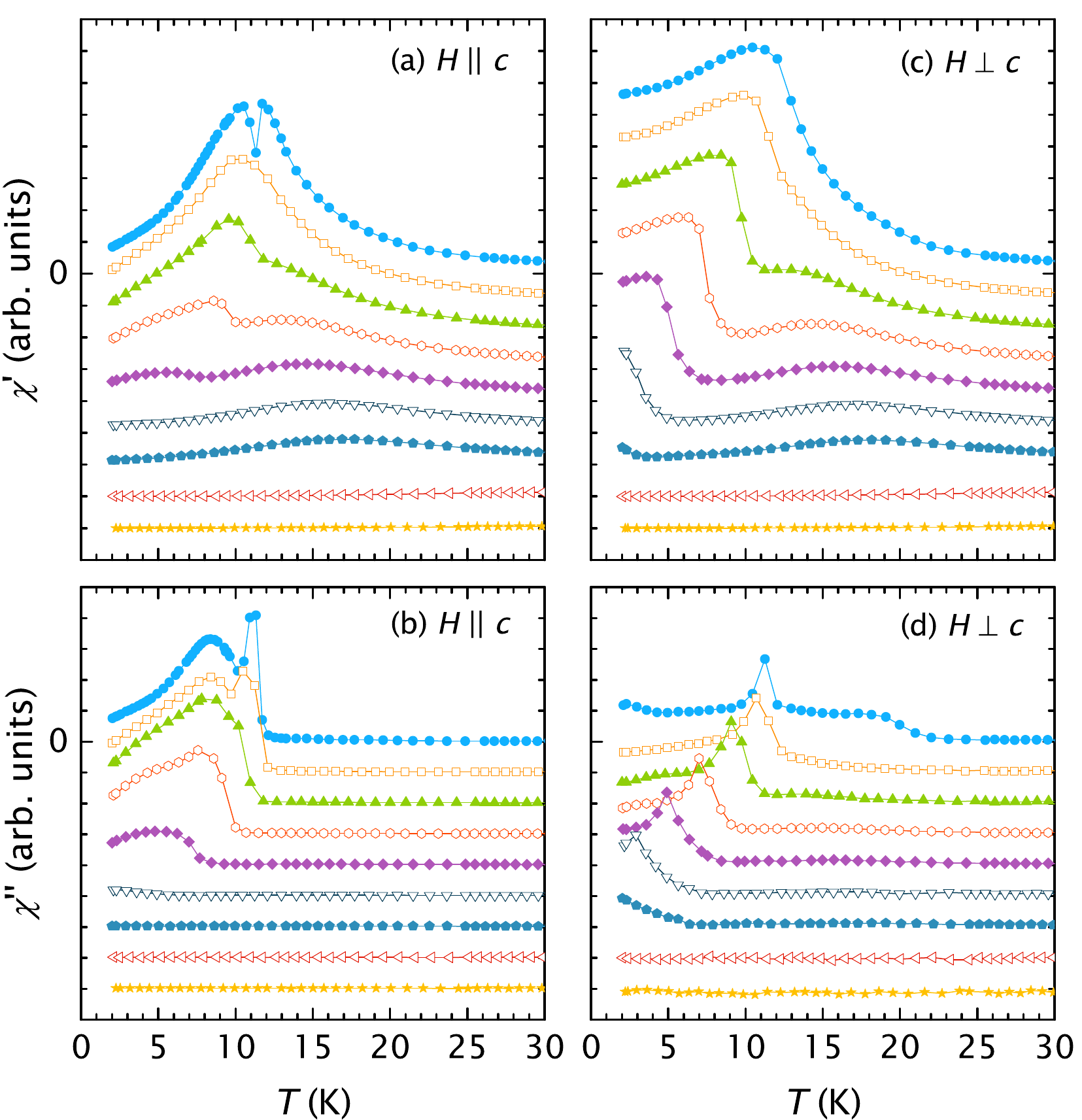}
    \caption{Temperature dependencies of (a, d) real $\chi'$ and (b, d) imaginary $\chi''$ parts of ac susceptibility investigated in (from top to bottom) 0, 0.1, 0.2, 0.3, 0.4, 0.5, 0.6, 5 and $\unit[9]{T}$ external magnetic fields applied (a, b) parallel and (c, d) perpendicular to the $c$-axis measured using a driving field of $\unit[10]{mT}$ with a frequency $\unit[1000]{Hz}$, data for subsequent fields are intentionally shifted along $y$-axis by $-1$ unit for better visibility \label{fig:ac}}
\end{figure}
\begin{figure}[h]
    \includegraphics[width=0.9\columnwidth]{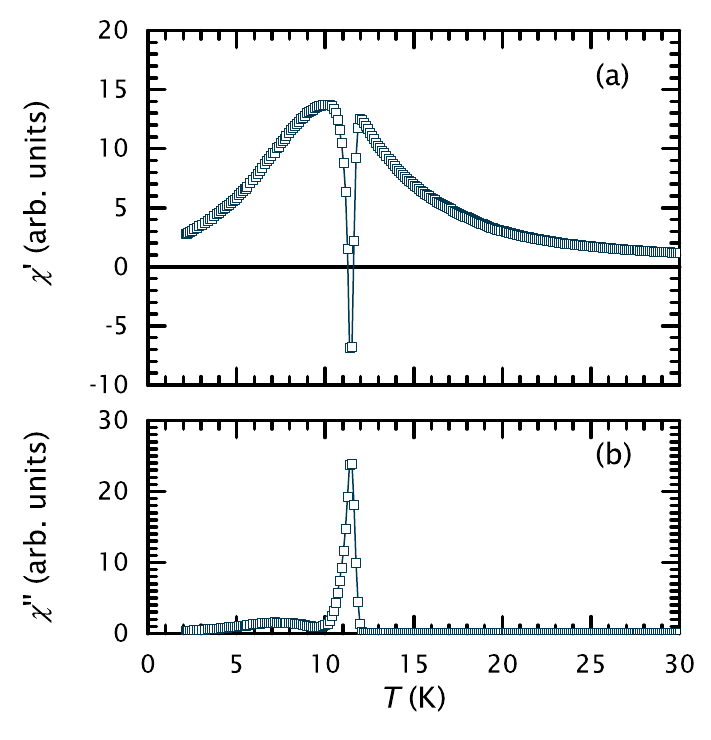}
    \caption{Temperature dependence of (a) real $\chi'$ and (b) imaginary $\chi''$ parts ac susceptibility measured in zero external magnetic field using a driving field of $\unit[10]{mT}$ with a frequency $\unit[1000]{Hz}$ applied parallel to the $c$-axis for another crystal of $\Euxii$ obtained from the same batch\label{fig:ac0}}
\end{figure}

The temperature dependence of ac susceptibility was measured in external magnetic fields applied either parallel or perpendicular to the \cax. The ac susceptibility measurements turned out to be a very sensitive technique, showing where most differences between individual crystals were observed. Nevertheless, the differences were in quantity not the quality of the results. In~Fig.~\ref{fig:ac} we present ac susceptibility measured for several external magnetic fields applied parallel and perpendicular to the \cax. For comparison in Fig.~\ref{fig:ac0} zero field ac susceptibility measured for another crystal (from the same batch) is shown.

The temperature dependent real part of susceptibility $\chi'(T)$ measured at zero-magnetic-field exhibits a broad maximum at around $\unit[12]{K}$ (cf. Fig. \ref{fig:ac} and \ref{fig:ac0}, see also Ref.~\citealp{Tran-SuppMat}).

For measurements with the driving field $\Hac$ applied parallel to the \cax, at $\Tc=\unit[12]{K}$, $\chi'(T)$ decreases rapidly and for some samples adopts negative values for a narrow 1-$\unit[2]{K}$ temperature range (cf. Fig.~\ref{fig:ac0}). At corresponding temperature the imaginary part of susceptibility $\chi''(T)$ has a sharp peak. Below $\unit[10]{K}$, $\chi''$ has another ``peak'', however a very broad one. With application of an external magnetic field, both the dip in $\chi'(T)$ and the sharp peak in $\chi''(T)$ shift to the lower temperatures with increasing magnetic field, additionally their intensities decrease and both the peak and the dip can not be distinguished above $\unit[0.25]{T}$. Additionally, two separate broad maxima in $\chi'(T)$ are observed, one of them shifts to the lower temperatures and the other to higher temperatures with increasing magnetic field. The broad ``peak'' in $\chi''(T)$ shifts to the lower temperatures and its intensity decreases at higher magnetic fields.

Similar features were observed for fields applied perpendicular to the \cax, i.e.,  (i) a sharp peak in the $\chi''(T)$ is visible at zero-magnetic-field at $\Tc=\unit[12]{K}$ and is shifted to lower temperatures with increasing magnetic field; (ii) there is also a wide maximum in $\chi'(T)$ at zero-magnetic-field that is divided into two separate maxima when the external magnetic field increases. However, there is no significant decrease of $\chi'$, i.e. there is no dip in $\chi'(T)$, at temperatures at which the $\chi''$-peak appears, as was observed for $H\parallel c$ measurements.

For both measurements \textendash{} in parallel and perpendicular field \textendash{} an additional broad maximum is observed in the real part of susceptibility, starting from around $\unit[12]{K}$ and shifts to higher temperatures with increasing magnetic field.

Comparing susceptibility measurements to the magnetoresistivity data, we associate the sharp peak in $\chi''(T)$ and the dip (or only change of slope) of $\chi'(T)$ with the appearance of superconductivity. The broad maximum starting at $\unit[12]{K}$, and shifting to the lower temperatures with increasing magnetic field, can be explained by the antiferromagnetic transition while the maximum that shifts to higher temperatures is most likely associated with the formation of the ferromagnetic component \textendash{} due to the reorientation of the $\Euion$ magnetic moments as a result of applied magnetic field, which is consistent with the magnetization measurements.

Based on these result transition temperatures were determined.

Using the Fisher's method, transition temperature to the antiferromagnetic order $\TN$ was determined from the real part of ac susceptibility. The $\TN(H)$ data points are summarized in Fig.~\ref{fig:Ph-diag} --- orange full circles. The $T$-$H$ points corresponding to the maxima in $\chi''(T)$ associated to the antiferromagnetic and superconducting transitions are summarized in Fig.~\ref{fig:Ph-diag} as orange open circles and full green triangles, respectively.

The transition line in the $T$-$H$ magnetic phase diagram between the paramagnetic (where the magnetic moments of $\Euion$ are paramagnetic) and antiferromagnetic states --- thus the $\Hcr^{*}$ field --- was determined as the high temperature maximum in the $\chi'(T)$.

\section{Discussion\label{sec:Discussion}}

\subsection{Magnetic phase diagram}

\begin{figure*}[!t]
    \includegraphics[width=1.9\columnwidth]{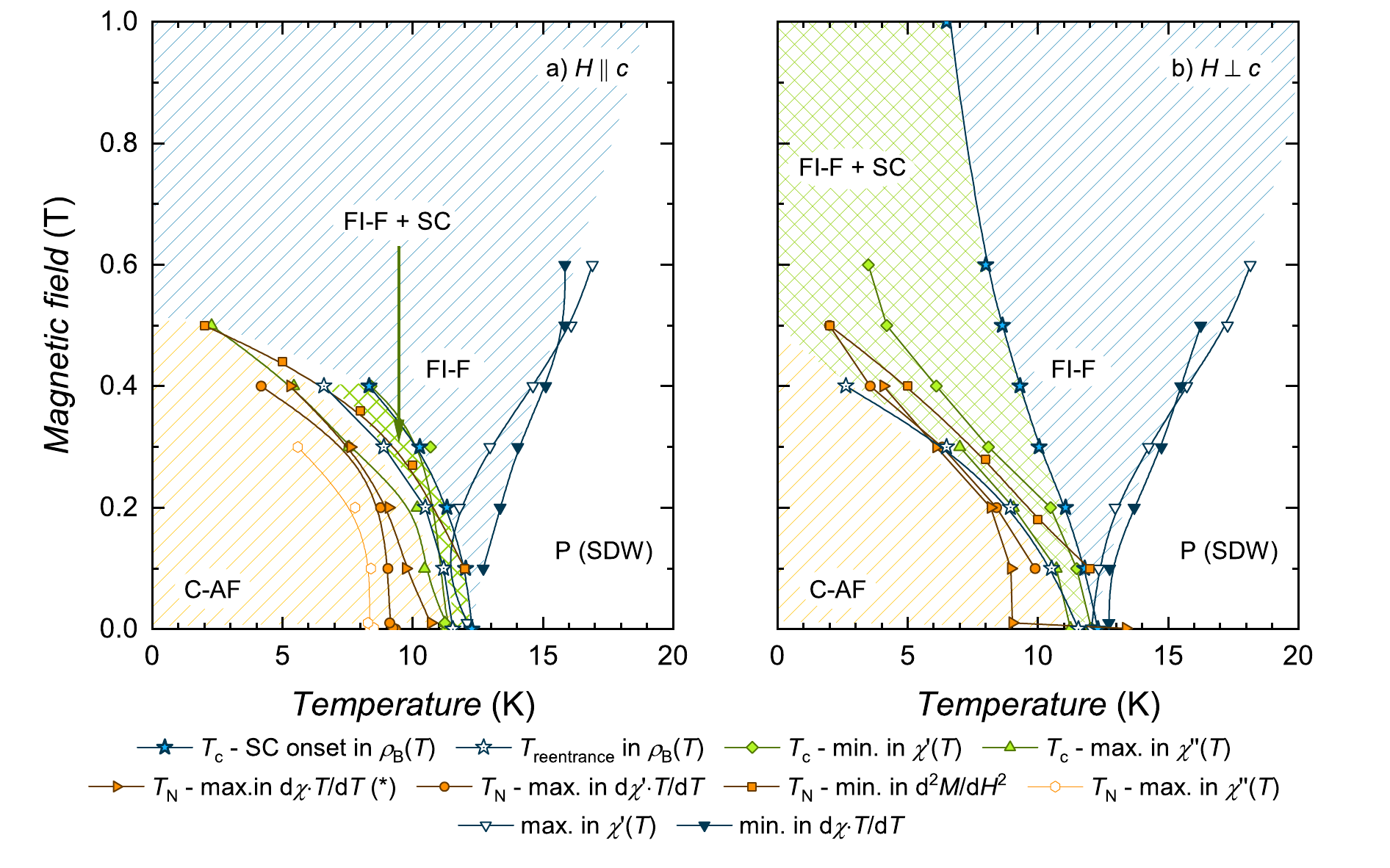}
    \caption{Magnetic phase diagram of $\Euxii$ for fields applied (a) parallel and (b) perpendicular to the $c$-axis\label{fig:Ph-diag}}
\end{figure*}

Based on the experimental results a magnetic phase diagram of $\Euxii$ was constructed and is presented in Fig.~\ref{fig:Ph-diag}.

For this system several magnetic phases can be distinguished: paramagnetic; paramagnetic with SDW order of $\Feion$ magnetic moments; superconducting; canted-antiferromagnetic (C-AF) and field induced ferromagnetic (FI-F) of $\Euion$ magnetic moments. Below $\TSDW=\unit[80]{K}$ the system exhibits a SDW order of $\Feion$ magnetic moments, therefore the system is truly paramagnetic only above this temperature and all other phases coexist with the SDW phase.

One should note, that although the field induced ferromagnetic state is present for fields applied parallel as well as for fields applied perpendicular to the crystallographic \cax, the microscopic order is remarkably different, i.e. for fields higher than $\Hcr$ (or $\Hcr^{*}$) the magnetic moment of $\Euion$ ions are aligned on the direction of the applied magnetic field --- for $H\parallel c$ the magnetic moments are parallel to the \cax{} and for $H\perp c$ the magnetic moments are perpendicular to the \cax.

\subsection{Superconductivity}

Although, as expected, by simultaneous Ca-doping the $\TN$ could be decreased and simultaneously $\Tc$ increased compared to the $\Euviii$, the zero-resistivity and diamagnetic signal are destroyed as soon as the Eu-magnetic-sublattice orders antiferromagnetically. This behavior is especially interesting when comparing presented results to the Co-doped systems with $0.075<x(\ce{Co})<0.2$, such as $\Euviii$, in which superconductivity appeared below $\TN$, therefore in an already antiferromagnetically ordered state \citep{Tran2012PRB,Tran2012NJP,Jin2016,Jin2013,Matusiak2011PRB,Ying2010PRB,Chen2010}. (It should be noted that in some papers the given compositions were based on the nominal Co concentrations, also EDX analysis is not the best tool for determining the concentration in case of Co and Fe compounds. It seems that the best way to determine Co concentration in the $\Eu$-based systems is by comparing the $\TSDW$ values.)

To explain why superconductivity can coexist with Eu-magnetic order in some of the Eu-122 systems (excluding $\Euxii$), we extend the previously proposed interpretation \citep{Tran2012PRB,Tran2012NJP}.

It was proposed by \citet{Klemm1975} that in layered superconductors the pairing between electrons (or holes) from different layers is minor. Therefore such systems will be anisotropic. The investigated superconductors have layered structure, where the FeAs-layers, responsible for superconductivity in these materials are perpendicular to the \cax, i.e., are spread on the \abplane. The \Eu-based systems are anisotropic, i.e. for \Euviii{} the anisotropy ratio was estimated to equal 2.4 \cite{Tran2012NJP}, while for \Euxii{} the anisotropy ratio $\Gamma = \frac{H_{|| ab}}{H_{\perp c}}$ is approximately $1.4$ at $T = \unit[8]{K}$ (see Supplementary Materials~\cite{Tran-SuppMat} for more details).

Superconductivity can be destroyed as a result of a paramagnetic pair breaking effect (PPB effect) or an orbital pair breaking effect (OPB effect).

While the PPB effect depends only on the intensity of the magnetic field (the greater the field, the greater the pair breaking), the contribution of the OPB effect in layered  superconductors with high spin-orbit scattering rate will also strongly depend on the direction of the magnetic field. In short, the OPB effect is not present when the magnetic field is parallel to the superconducting layers, on the other hand when the field is applied perpendicular to the superconducting layers the OPB effect is maximal.

The Co-doped $\Eu$-based systems are layered superconductors with a high spin-orbit scattering rate, however due to the magnetic moment of $\Euion$ even with no external magnetic field applied, the superconducting layers are exposed to a magnetic field. Since the magnetic moments are canted from the \abplane, there are two magnetic components: parallel and perpendicular to the superconducting layers. Depending on the composition of the compound, the contribution of the components will be different.

When comparing $\Euxii$ and $\Euviii$ (or other Eu-122 systems with $0.075<x(\ce{Co})<0.2$) one has to have in mind that the microscopic magnetic structure on Eu-sublattice is different for these two compounds. As shown by the \MS{} and neutron scattering, the angle between the magnetic moment of Eu and the \cax{} $\vartheta$ changes with Co-doping \citep{Blachowski2011Codoped,Jin2016}. Although for both $\Euxii$ and $\Euviii$ their ground state is canted antiferromagnetic (C-AF), based on \MS{} study, it is known that  in zero-magnetic field the angle $\vartheta$ is smaller for the Ca-and-Co-doped compound ($\vartheta\in\left\langle 44^{\circ},60^{\circ}\right\rangle $) than for the Co-only-doped one with $\vartheta\approx30^{\circ}$ \citep{Komedera2018,Tran2017,Blachowski2011Codoped}. In other words in the ground state, the superconducting layers  experience a perpendicularly applied magnetic field (due to the magnetic moment of $\Euion$), however it is bigger for the $\Euxii$ than in the case of the $\Euviii$ compound. This is probably the reason why superconductivity at low temperatures is destroyed for the Ca-and-Co-doped compound while not for Co-doped one.

By applying an external magnetic field the magnetic moments of $\Euion$ can be ``rotated'' with the final form of an field induced ferromagnetic (FI-F) order. Here again one should note that the FI-F order is contrasting between the situations at which the field is applied parallel and perpendicular to the \cax{}. Consequently, when the field is applied perpendicular to the \cax{} the OPB effect is minimal, both because the external magnetic field and the magnetic moments of $\Euion$ are parallel to the superconducting layers, and so superconductivity is mostly destroyed due to the PPB effect. On the contrary, the OPB effect is maximal when the field is parallel to the \cax{}, especially above $\Hcr$, when also the magnetic moments of $\Euion$ are perpendicular to the superconducting layers.

\section{Summary}

We have investigated magnetic and electrical transport properties of $\Euxii$ compound and compared its properties to similar members of the $\Eu$-based family. Depending on the temperature and magnetic field applied, $\Euxii$ undergoes multiple phase transitions. With decreasing the temperature we observe a transition from (i) paramagnetic state to (ii) spin density wave order of $\Feion$ followed by (iii) superconducting transition and (iv) canted-antiferromagnetic order of $\Euion$ magnetic moments. With application of an external magnetic field
the $\Euion$ magnetic moment reorient and
(v) field-induced ferromagnetic order
can also be observed. For graphical presentation a magnetic phase diagram was constructed (see Fig.~\ref{fig:Ph-diag}). Depending on the direction of applied external magnetic field, superconductivity can coexist with the $\Euion$ magnetic moments order.

The parameters calculated from electrical transport data are comparable to those obtained for other Co-doped $\Eu$-based compounds.

By simultaneous Co and Ca-doping we were able to obtain a compound with superconductivity appearing above the antiferromagnetic order of $\Euion$ magnetic moments. However, a re-entrance behavior was observed --- instead of zero resistivity and diamagnetic signal down to the lowest temperatures. We show that this behavior is a result of simultaneous coincidence of few effects --- the anisotropy expected for layered superconductors~\citep{Klemm1975}, the difference of contributions of orbital pair breaking effect depending on the direction of the magnetic field in relation to the superconducting pairs in such systems and the magnetic structure of the compound, that also depends on the external magnetic field.

\bibliography{Eu12publi_1-1}

\end{document}


\title{Supplementary Materials: Re-entrance of resistivity due to the interplay
\\
of superconductivity and magnetism\\
in \ce{Eu_{0.73}Ca_{0.27}(Fe_{0.87}Co_{0.13})2As2}}
\author{Lan Maria Tran}
\email[Corresponding author: ]{l.m.tran@intibs.pl}

\affiliation{Institute of Low Temperature and Structure Research, Polish Academy of Sciences}
\author{Andrzej J. Zaleski}
\affiliation{Institute of Low Temperature and Structure Research, Polish Academy of Sciences}
\author{\framebox{Zbigniew Bukowski}}
\affiliation{Institute of Low Temperature and Structure Research, Polish Academy of Sciences}
\date{\today}

\maketitle

\section{Repeatibility}

For this study multiple crystals of $\Euxii$ grown from the same
batch were measured.

The composition of the material was determined using EDX analysis.
It should be noted, that while the accuracy of determining the composition
by EDX is of $1\%$, in the specific case of compounds containing
Fe and Co, the accuracy is below approximately $5\%$. This is caused
by the small differences of the energy between Fe and Co spectra and
the overlapping of the intensities belonging to the separate elements.
The composition of $\Euxii$ was determined by taking the average
composition measured at several points for several crystals. The difference
of composition for these points was less than $5\%$, therefore was
smaller than the EDX resolution.

Although according to EDX analysis the samples had more or less the
same compositions, the properties moderately varied between crystals:
(i) we observed different critical temperatures for subsequent transitions
and (ii) the impact of the phase transition on observed physical properties
were different for different samples (e.g. the resistivity drop at
$\Tc$ was not always to zero). Nevertheless, we can show that qualitatively
the samples are the same, and observed properties can be explained
with the same physical model. Consequently, even small differences
in concentrations distribution (not detectable by e.g. EDX) can have
a big impact on the properties of the compound. To demonstrate these
conclusions we show some example results.

\begin{figure}[h] \includegraphics[width=0.99\columnwidth]{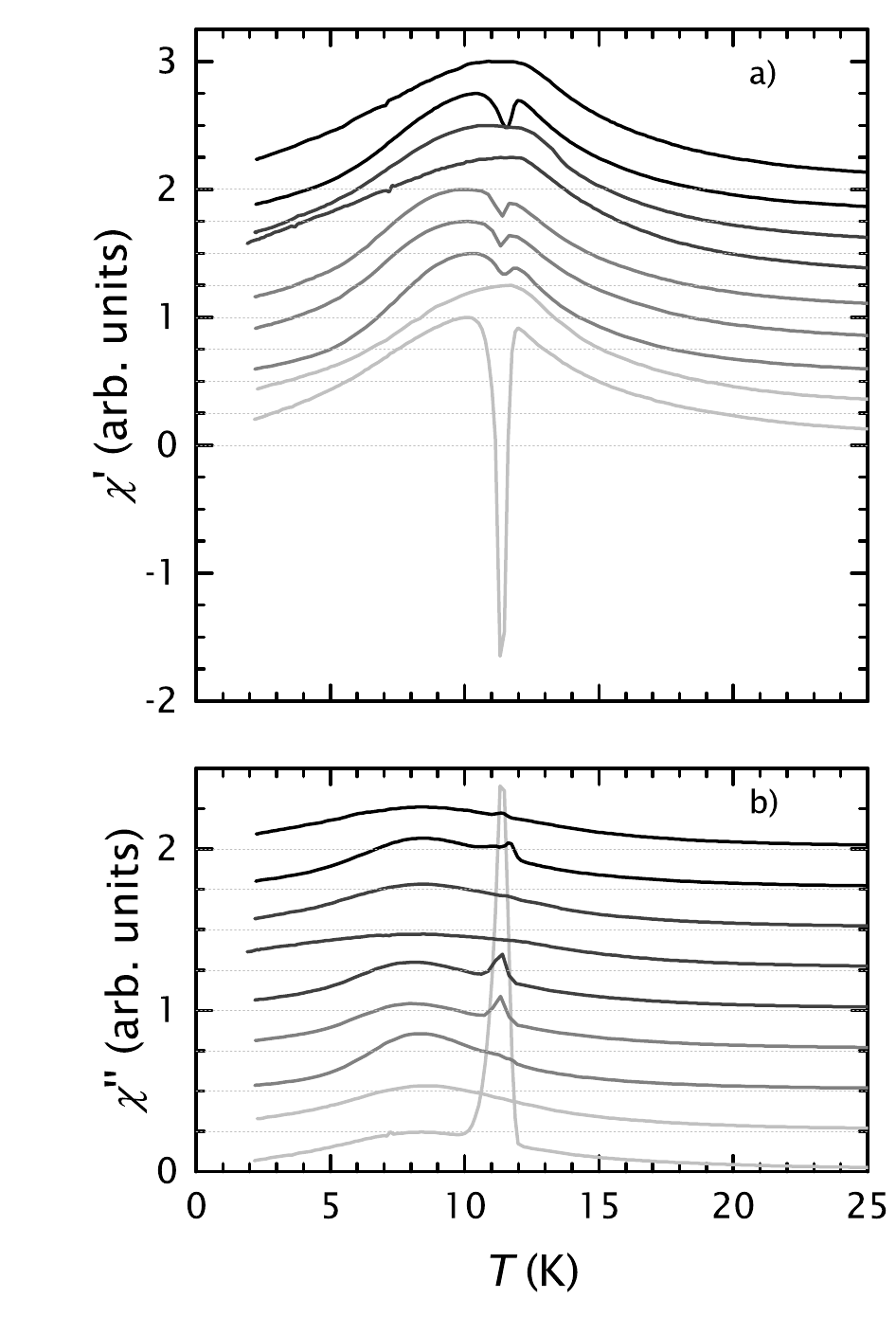}\caption{Temperature dependence of normalized ac susceptibility for several crystals from the same batch investigated in zero magnetic field using a $\unit[10]{mT}$ driving field ($\Hac\parallel c$) with a frequency of $\unit[1000]{Hz}$; the data for subsequent samples are shifted by $0.25$ units along Y-axis for better visibility, the dashed lines denote 0 for each set of data \label{fig:Chi}}
\end{figure}
For comparison, temperature dependencies of zero field ac susceptibility
of several crystals grown from the same batch are presented in Fig.~\ref{fig:Chi}.
As can be seen the quality of the data between the samples is comparable,
i.e. all samples exhibit a broad maximum around $\unit[12]{K}$ and
a drop of the real part of susceptibility and a sharp peak of the
imaginary part below $\unit[12]{K}$ (cf.~Fig.~\ref{fig:Chi}).
The transition temperatures shift by up to $\unit[2]{K}$ and the
intensity of the drop and peaks vary between the samples. Some differences
in behavior \textemdash{} smaller or bigger ``dip'' due to the superconductivity,
corresponding to the volume of the superconducting phase \textemdash{}
suggest that the differences in concentrations distribution between
crystals is enough to change the volume of superconducting phase.

\begin{figure*}[t] 
    \includegraphics[width=1.98\columnwidth]{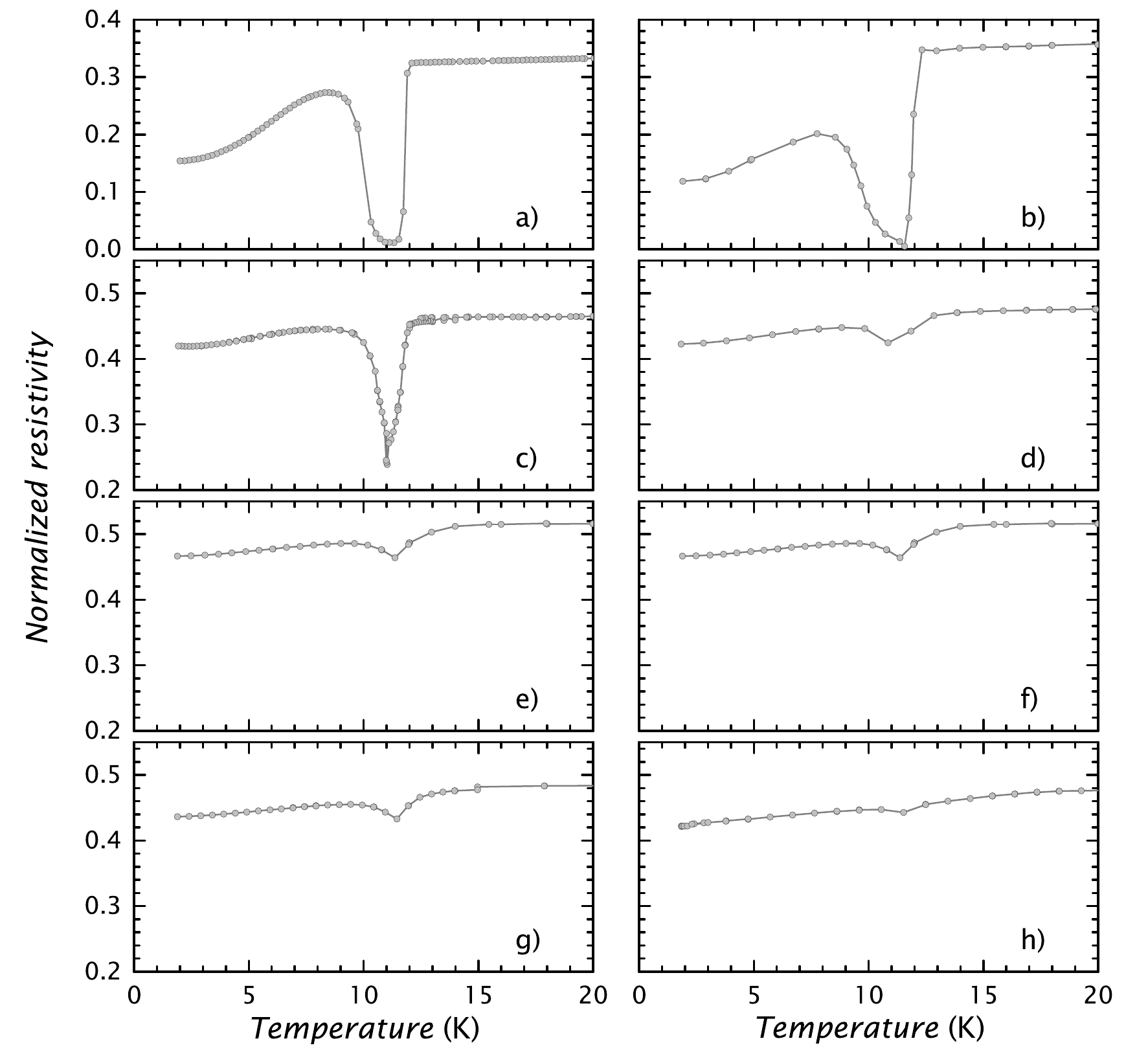}
    \caption{Temperature dependence of resistivity normalized to resistivity at $\unit[300]{K}$ for several crystals obtained from the same batch \label{fig:Res}}
\end{figure*}

Similarly, in Fig.~\ref{fig:Res} temperature dependencies of resistivity
for chosen crystals are shown. The somewhat sharp drop and the reentrance
of resistivity is present in all samples, however depending on the
sample the resistivity does not always drop to zero. Some differences
of the onset and width of transition are also observed.

Although, some quantitive differences of investigated physical properties
are observed, the reentrance behavior of ac susceptibility and resistivity
are detected for all samples. It should be noted however, that the
presence of superconductivity and this reentrance behavior can be
changed by just minor change in the concentrations, smaller than the
experimental error from the EDX measurements for this group of compounds.

\section{Transition temperature determination}

\begin{figure*}[t] 
    \includegraphics[width=1.9\columnwidth]{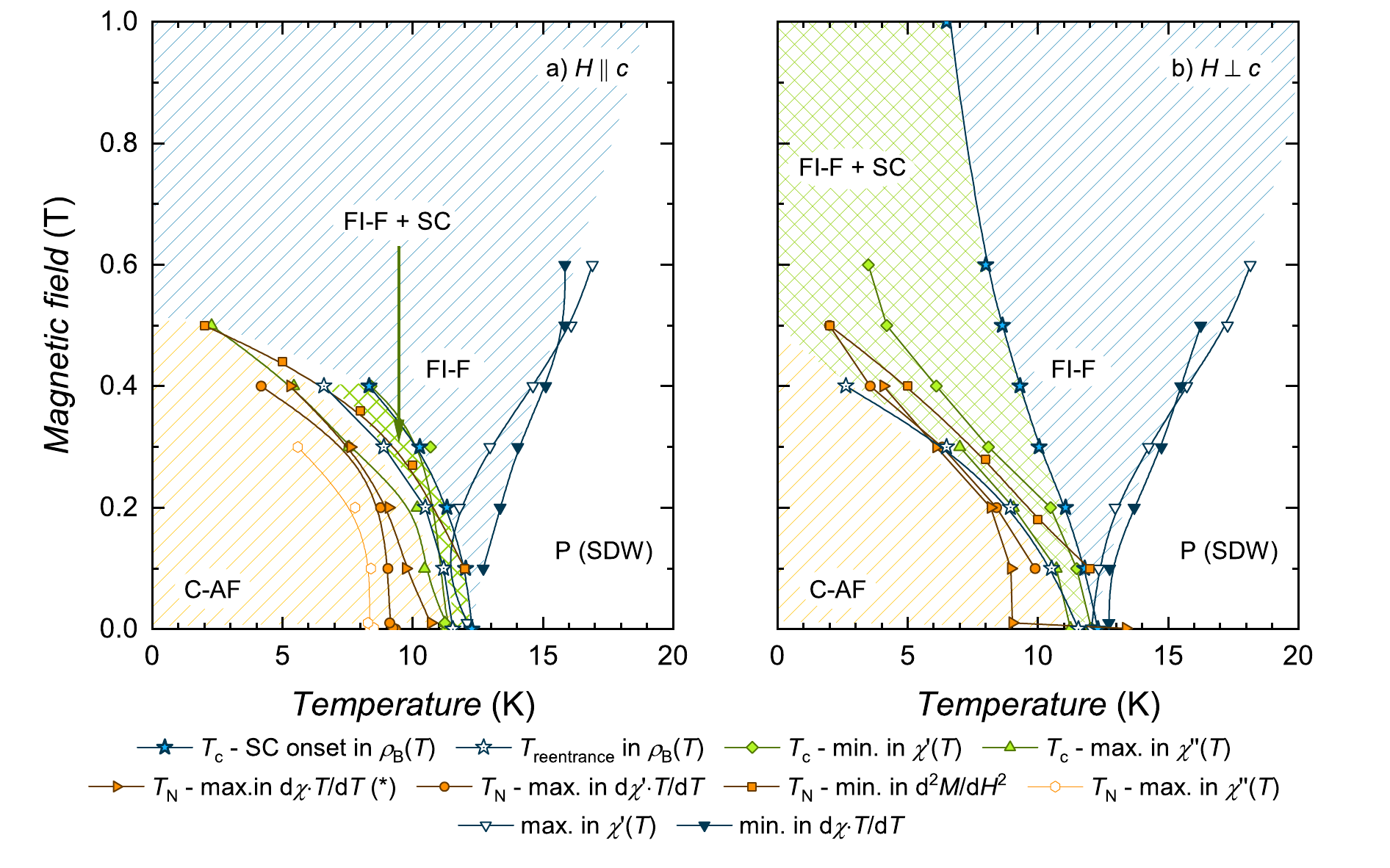}
    \caption{Magnetic phase diagram of $\Euxii$ for fields applied (a) parallel and (b) perpendicular to the $c$-axis\label{fig:Ph-diag}}
\end{figure*}
\begin{figure}[h] 
    \includegraphics[width=0.99\columnwidth]{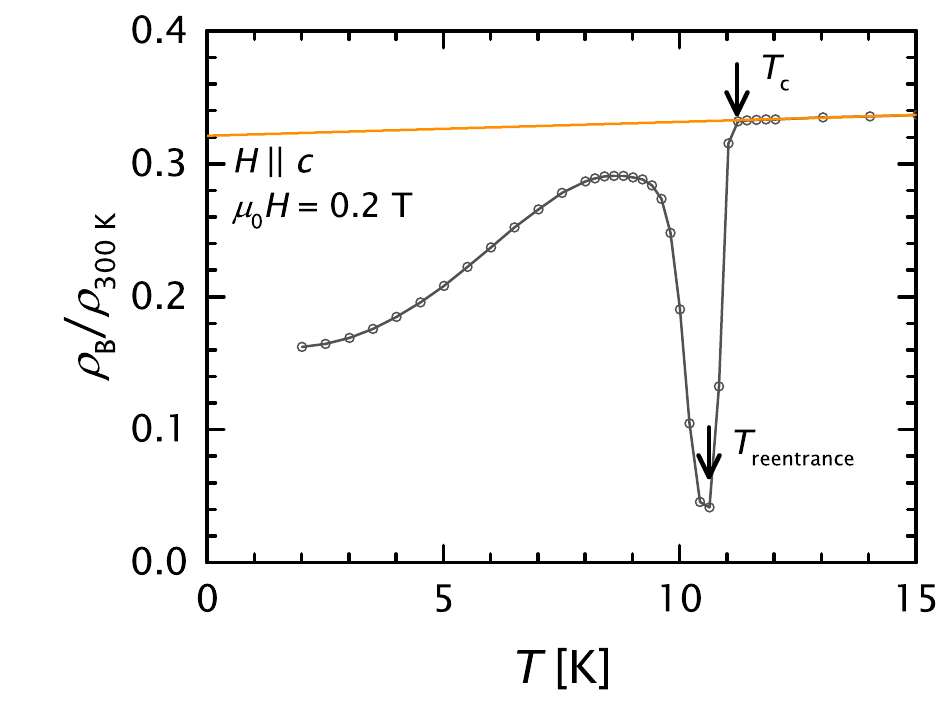}
    \caption{Temperature dependence of resistivity investigated in $\unit[0.2]{T}$ magnetic field applied parallel to the $c$-axis: Example of how the temperature of the superconducting transition and reentrance was determined from resistivity and magnetoresistivity\label{fig:TC-Res}}
\end{figure}
\begin{figure}[h] 
    \includegraphics[width=0.99\columnwidth]{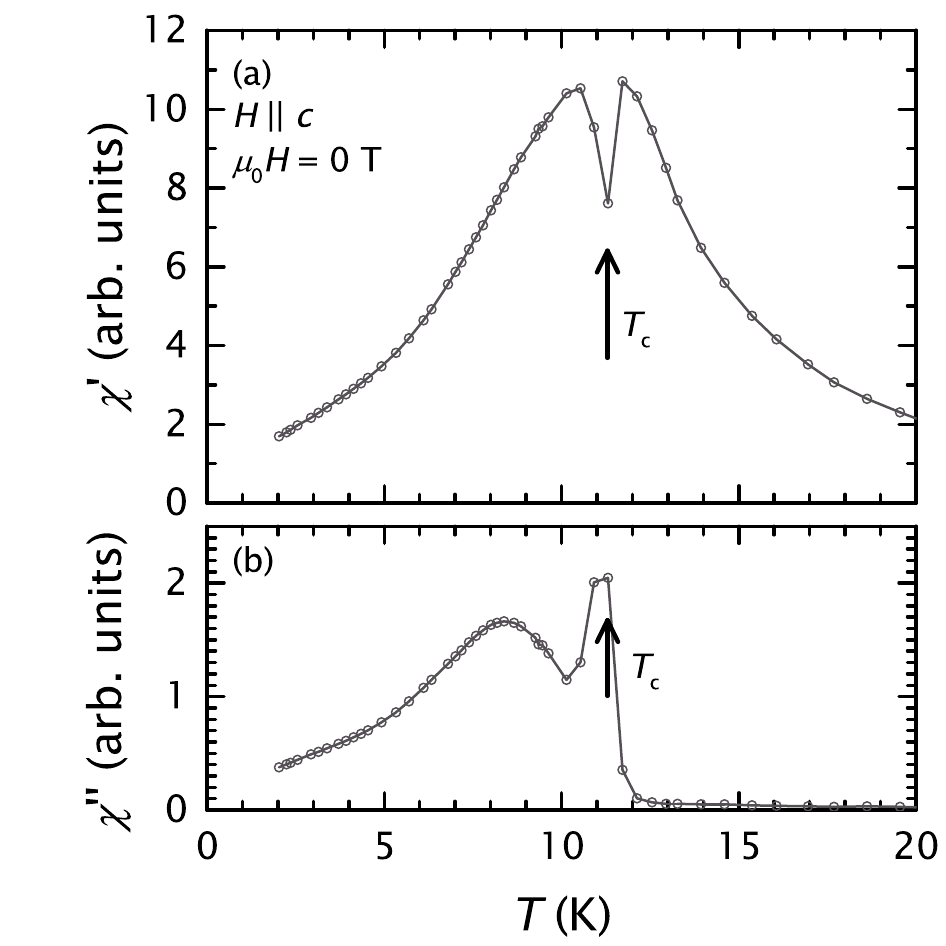}
    \caption{Temperature dependence of ac (a) real and (b) imaginary parts of magnetic susceptibility investigated using $\Hac=\unit[10]{mT}$ driving field with $f=\unit[1]{kHz}$ frequency at zero external magnetic field: Example of how the temperature of superconducting transition was determined from ac susceptibility measurements\label{fig:TC-Susc}}
\end{figure}
\begin{figure}[h] 
    \includegraphics[width=0.99\columnwidth]{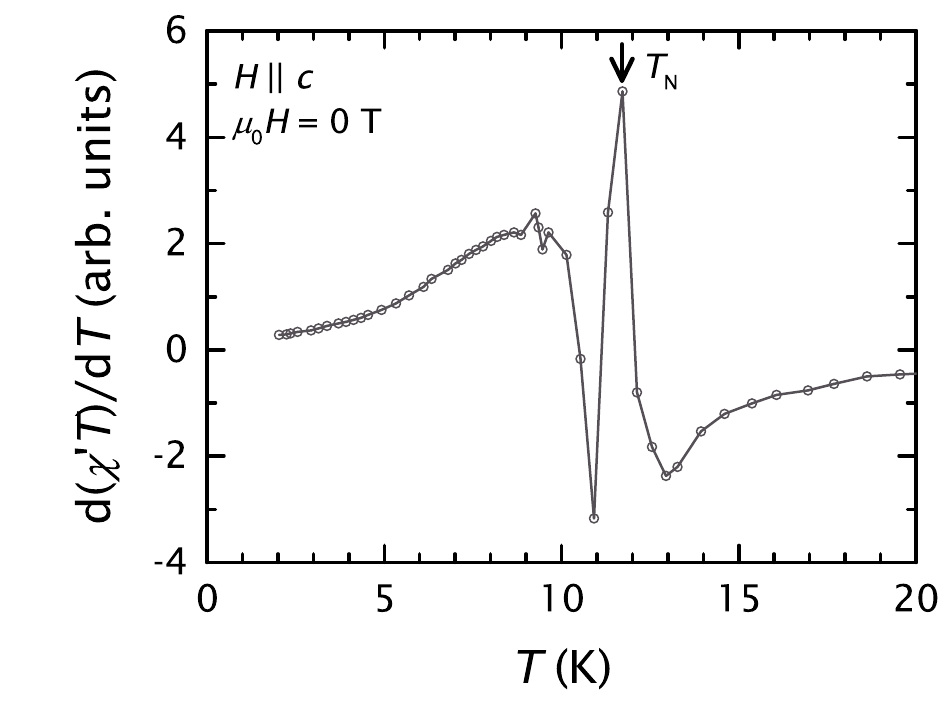}
    \caption{Temperature dependence of first derivative of product of real part of ac susceptibility and temperature $\nicefrac{\dd\chi'T}{\dd T}$ for measurements using driving field $\Hac=\unit[1]{mT}$ applied parallel to the $c$-axis with frequency $f=\unit[1]{kHz}$ investigated in zero magnetic field: Example of how the Ne\'{e}l temperature was determined from ac susceptibility measured in magnetic fields applied parallel and perpendicular to the $c$-axis, and from susceptibility $\chi=\nicefrac{M}{H}$ investigated in fields perpendicular to the $c$-axis \label{fig:Fisher}}
\end{figure}
\begin{figure}[h] 
    \includegraphics[width=0.99\columnwidth]{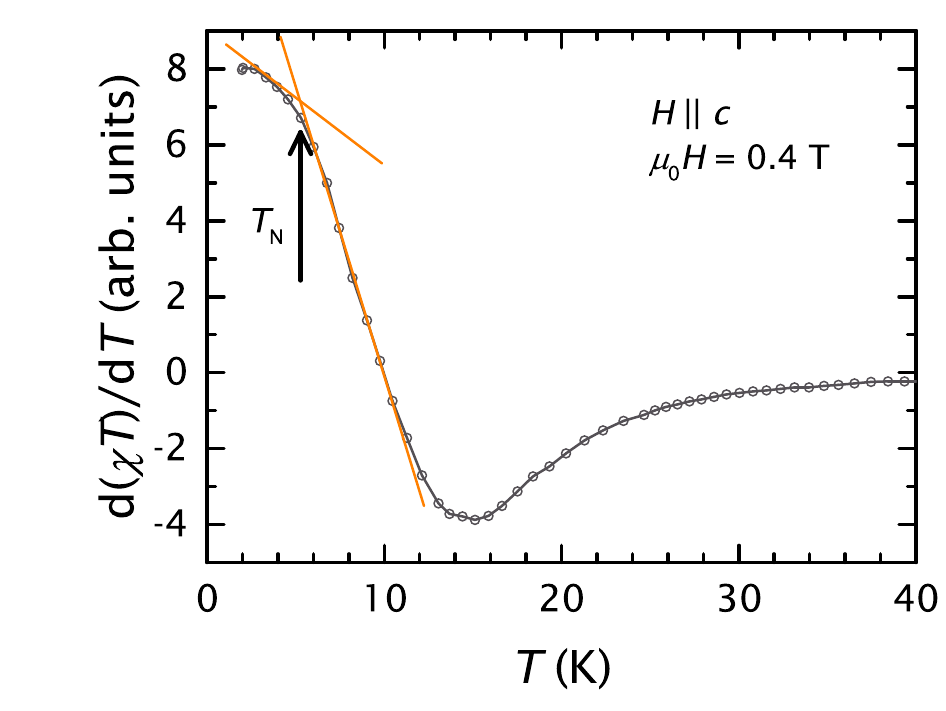}
    \caption{Temperature dependence of first derivative of product of susceptibility and temperature $\nicefrac{\dd\chi T}{\dd T}$ for measurements in $\unit[0.4]{T}$ magnetic field applied parallel to the $c$-axis: Example of how the Ne\'{e}l temperature was determined from susceptibility $\chi=\nicefrac{M}{H}$ investigated in magnetic fields applied parallel to the $c$-axis \label{fig:Fisher-Hc}}
\end{figure}
\begin{figure}[h] 
    \includegraphics[width=0.99\columnwidth]{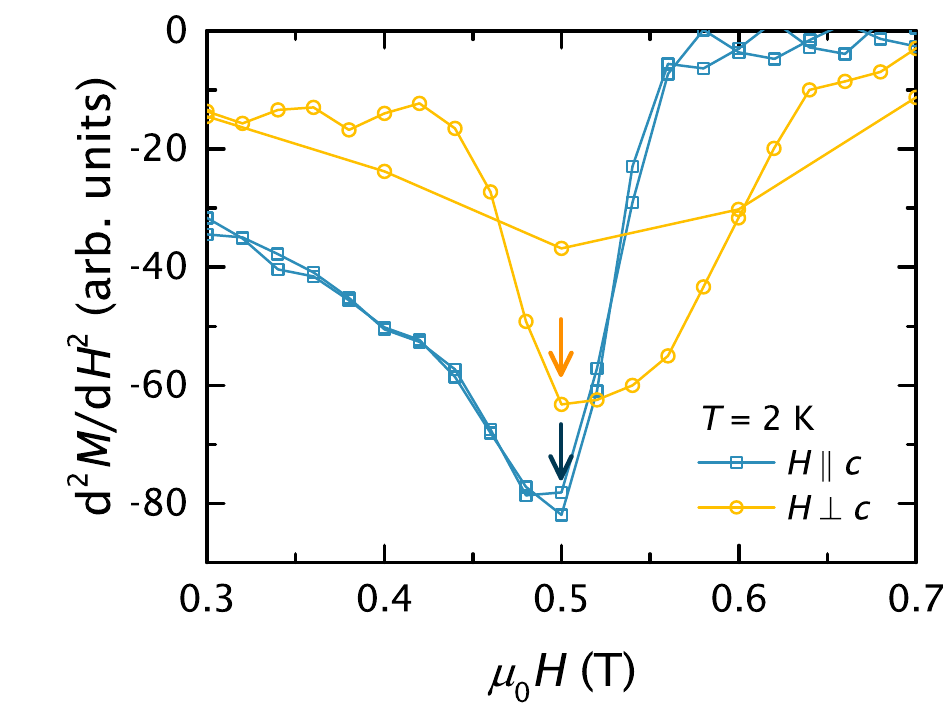}
    \caption{Magnetic field dependence of second derivative of magnetization for measurements at $\unit[2]{K}$ for fields applied parallel and perpendicular to the $c$-axis: Example of how the crossover field $\Hcr$ was determined from magnetization measurements \label{fig:TC-Mag}}
\end{figure}

Based on the experimental results a magnetic phase diagram of $\Euxii$
was constructed and is presented in Fig.~\ref{fig:Ph-diag}.

The data points were determined as follow. From magnetoresistivity
measurements the critical temperature $\Tc$ and reentrance temperature
were estimated. The $\Tc$ (full blue stars in Fig.~\ref{fig:Ph-diag})
was determined as the temperature at which the resistivity (or magnetoresistivity)
$\rho_{\text{B}}(T)$ starts to decrease (cf. ``$\Tc$'' arrow in
Fig.~\ref{fig:TC-Res}). The reentrance temperature (open blue stars
in Fig.~\ref{fig:Ph-diag}) are the temperatures at which the resistivity
starts to increase below the superconducting transition (cf. ``$T_{\text{reentrance}}$''
in Fig.~\ref{fig:TC-Res}). Additionally, from susceptibility measurements
$\Tc$ was estimated as the local minimum in the real part of susceptibility
$\chi'$ \textemdash{} cf. ``$\Tc$'' arrow in Fig.~\ref{fig:TC-Susc}(a)
\textemdash{} and as the maximum in the imaginary part of susceptibility
\textemdash{} cf. ``$\Tc$'' arrow in Fig.~\ref{fig:TC-Susc}(b);
the respective data points are presented as full green diamonds in
Fig.~\ref{fig:Ph-diag} and full green triangles in Fig.~\ref{fig:Ph-diag}.

The temperature transition to the antiferromagnetic state and the
field-induced ferromagnetic state were determined from magnetization
and susceptibility measurements. One way to determine the Ne\'el
temperature is by using the Fisher's method \citep{Fisher1962}. For
this method the temperature dependencies of $\dd\chi(T)T/\dd T$,
where $\chi$ can be either the real part of susceptibility $\chi'$
or susceptibility calculated from magnetization $\chi=M/H$, were
determined. The maximum of this function is at Ne\'{e}l temperature
(cf. ``$\TN$'' arrow in Fig.~\ref{fig:Fisher}). A clear maximum
in the temperature dependence of $\dd\chi(T)T/\dd T$ was observed
for the real part of susceptibility measured in both fields applied
parallel and perpendicular to the \cax; collected data points are
presented as full circles in Fig.~\ref{fig:Ph-diag}. Also for susceptibility
calculated from magnetization measurements in magnetic fields applied
perpendicular to the \cax, Fisher's method could be easily applied.
However, for the measurements in fields applied parallel to the \cax
the $\nicefrac{\dd\chi(T)T}{\dd T}$ vs. $T$ function does not have
a clear maximum but rather increases down to lowest investigated temperatures,
see Fig.~\ref{fig:Fisher-Hc}. In this case the $\TN$ was estimated
as the temperature at which the linear extrapolations of two regions
cross. The data points evaluated from $\chi(T)$ are presented as
full orange triangles in Fig.~\ref{fig:Ph-diag}. The imaginary part
of susceptibility for measurements in magnetic field applied parallel
to the \cax has a vivid maximum associated with the antiferromagnetic
transition, the temperature at which the maximum is reached was selected
as the $\TN$ (collected as open orange circles in Fig.~\ref{fig:Ph-diag}),
cf. ``$\TN$'' arrow in Fig.~\ref{fig:TC-Susc}(b).

The transition from the antiferromagnetic to field induced ferromagnetic
state is clearly observed in the field dependence magnetization data
as the point at which the $M(H)$ dependence saturates, the ``crossover''
field $\Hcr$ was determined from the minimum in the field dependence
of second derivative of magnetization $\dd^{2}M/\dd H^{2}$ (cf. Fig.~\ref{fig:TC-Mag}),
the data points are collected as full orange squares in Fig.~\ref{fig:Ph-diag}.

\section{Upper critical field}

\begin{figure}[h] 
    \includegraphics[width=0.99\columnwidth]{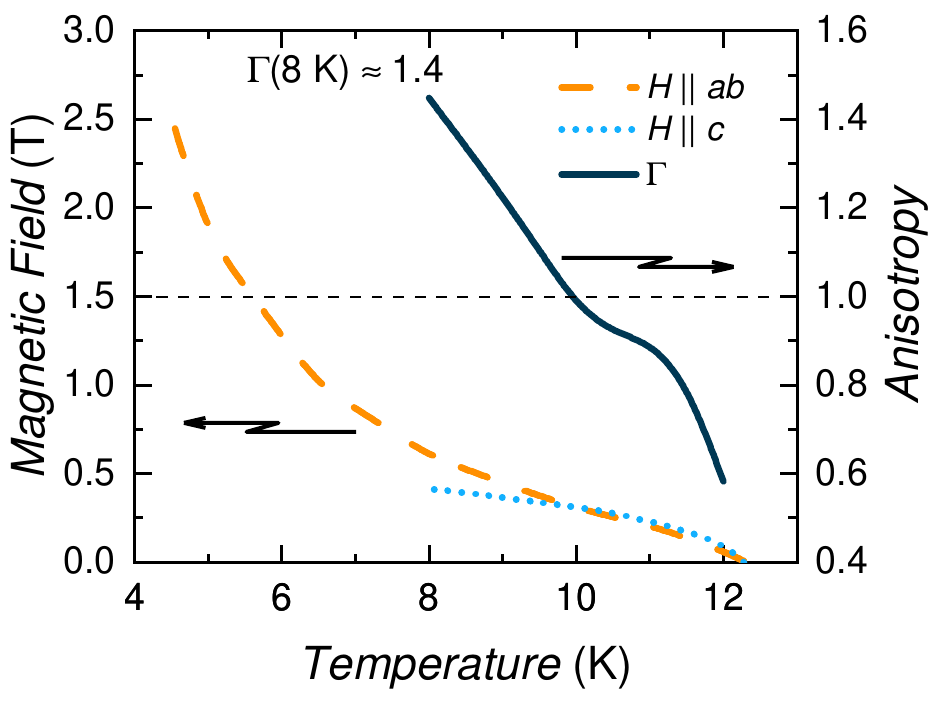}
    \caption{Extrapolation of $\Tc$ - SC onset in $\rho_B(T)$ from Fig.~\ref{fig:Ph-diag} for magnetic field applied parallel (dotted blue line) and perpendicular (dashed orange line) to the $c$-axis\label{fig:Anisotropy}}
\end{figure}

Taking extrapolation of $H_\text{c}(\Tc)$ (based on the $\Tc$ onset in $\rho_B(T)$ points in Fig.~\ref{fig:Ph-diag}) the upper critical fields for fields applied parallel and perpendicular to the \cax{} can be compared. The determined dependencies are presented in Fig.~\ref{fig:Anisotropy} along with anisotropy of the upper critical field $\Gamma$ calculated as 
\begin{equation}
    \Gamma=\dfrac{\Hc^{||ab}}{\Hc^{||c}}.
\end{equation}
As can be seen, at $\unit[8]{K}$, the lowest temperature the upper critical fields can be compared, anisotropy $\Gamma$ is approximately $1.4$. 


\section{Induced superconductivity}

As implied in the main paper \citep{Tran-thisPaper}, superconductivity
can coexist with the magnetic order as long as the magnetic component
parallel to the \cax{} is minor or none. Here we present some additional
figures with schematic explanation on this problem.

\begin{figure*}[p] 
    \includegraphics[width=0.95\columnwidth]{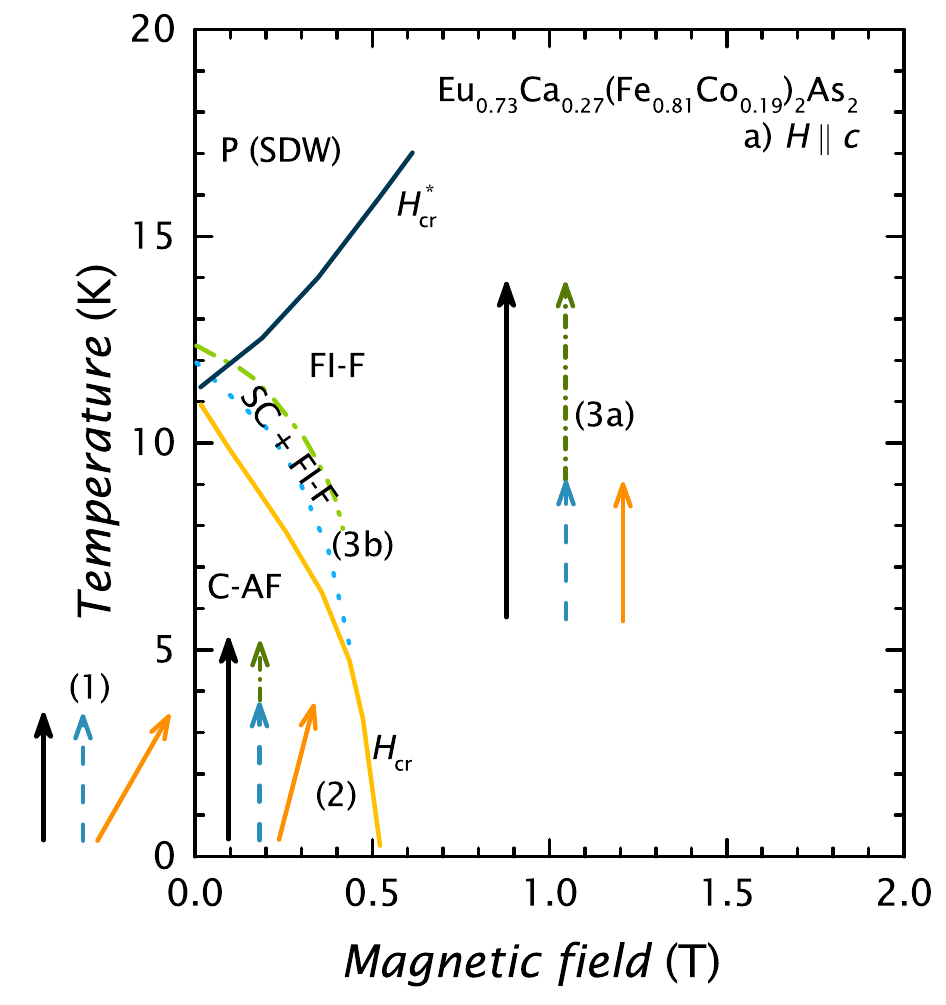}\hspace{15pt}
    \includegraphics[width=0.95\columnwidth]{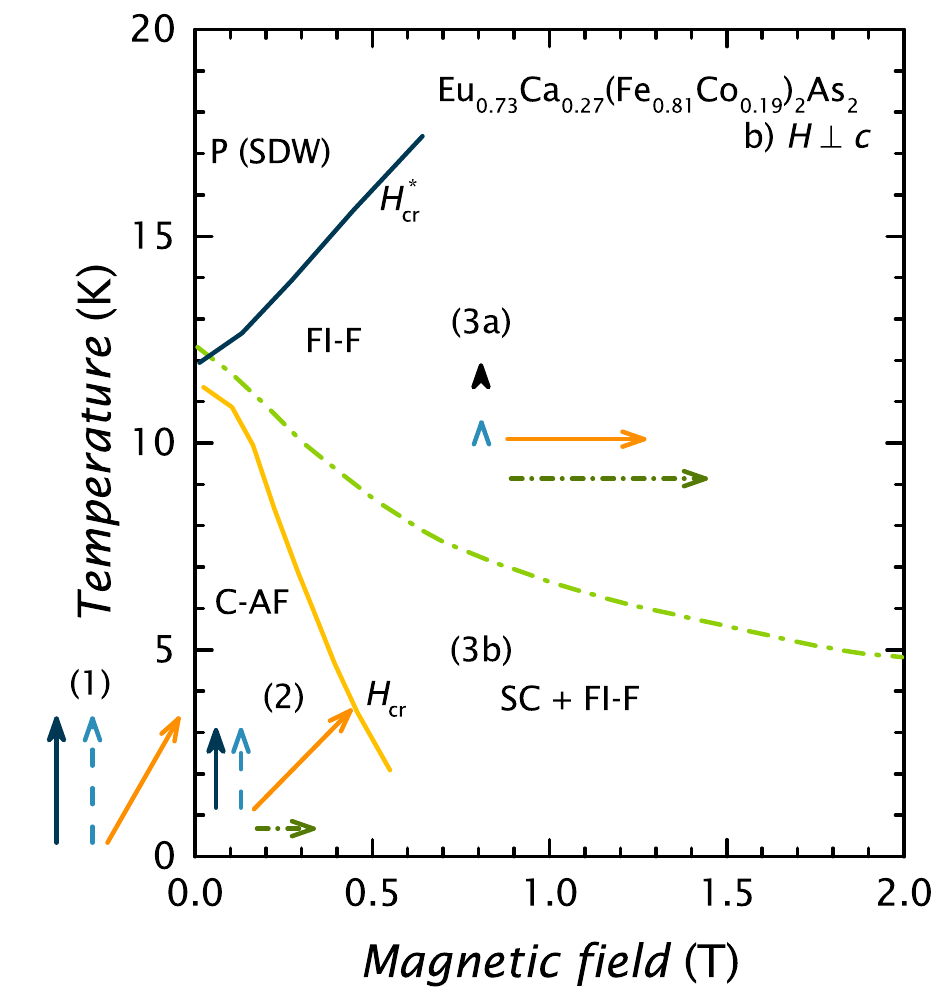}
    \includegraphics[width=0.95\columnwidth]{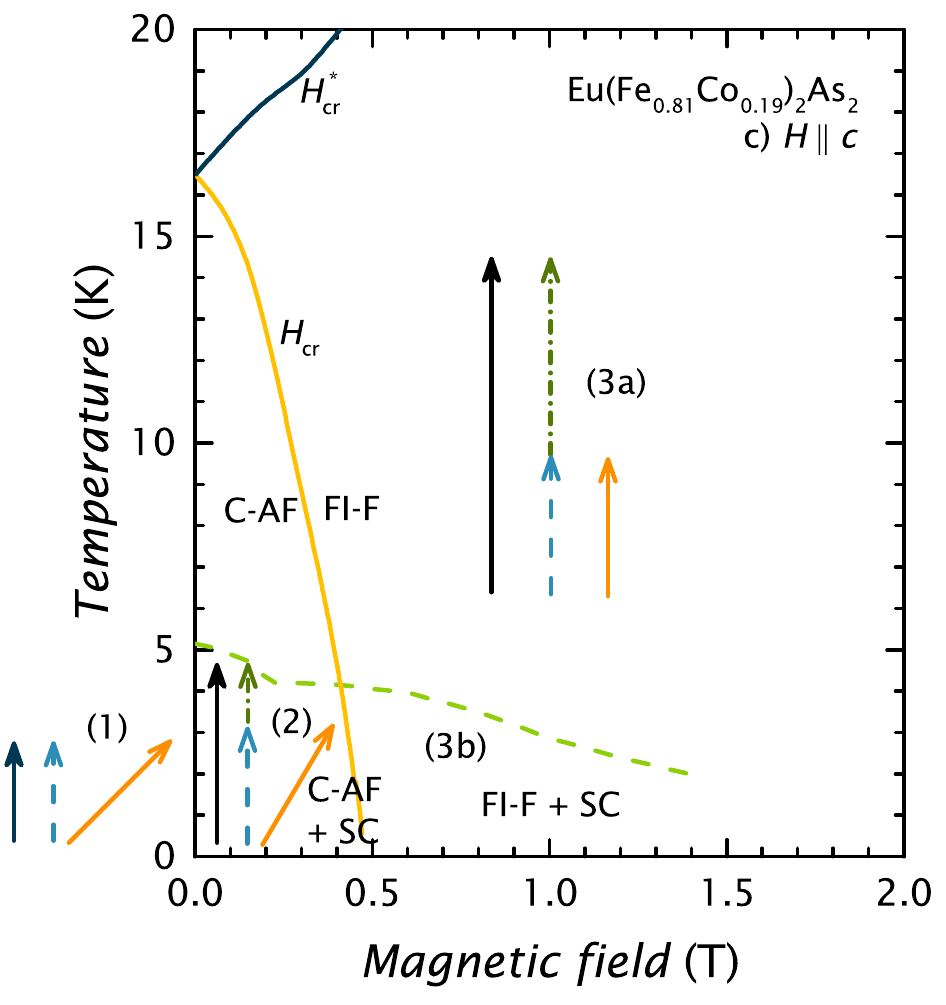}\hspace{15pt}
    \includegraphics[width=0.95\columnwidth]{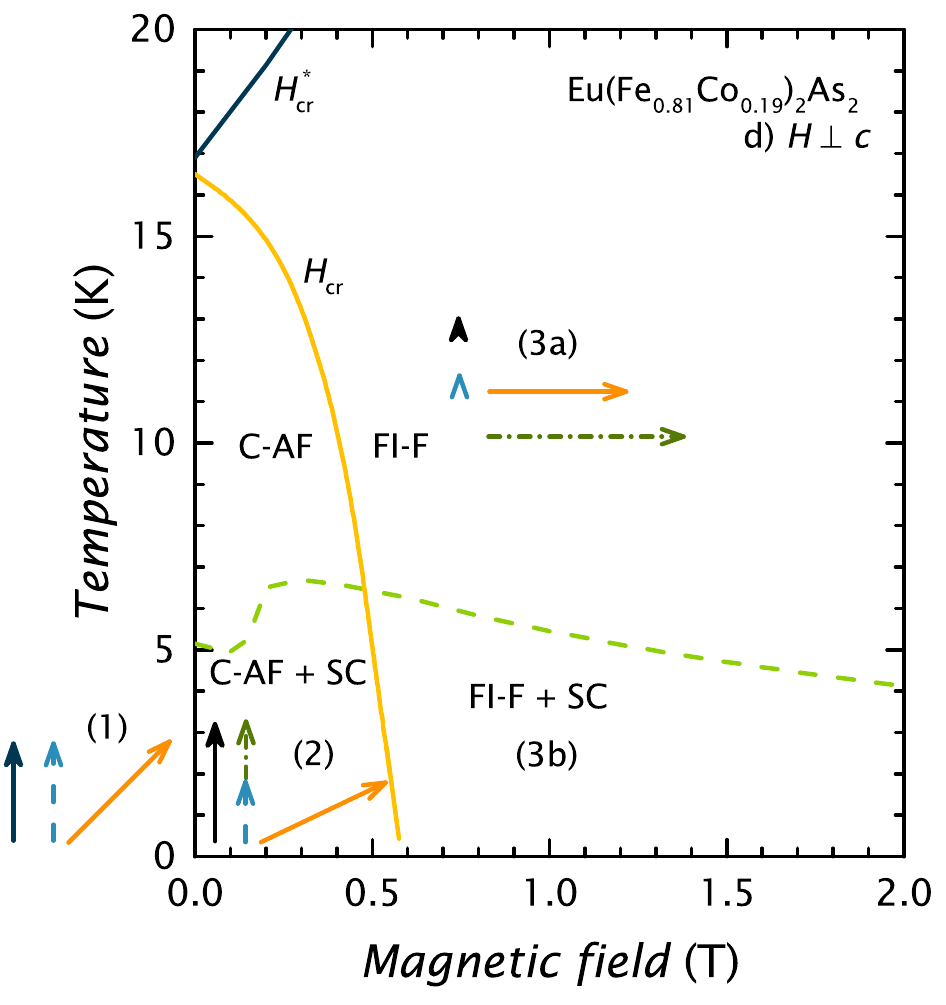}
    \caption{Simplified magnetic phase diagram of (a\textendash b) $\Euviii$ and (c\textendash d) $\Euxii$ with graphical illustration of the change of direction of magnetic moment of $\Euion$ (orange arrow) and the respective component parallel to the $c$-axis (blue dashed arrow) in dependence of the external magnetic field (schematically presented as dash-doted green arrow) applied either (a, c) parallel or (b, d) perpendicular to the $c$-axis, the dark solid arrow represents the summary component of the magnetic field parallel to the $c$-axis \label{fig:Diag}}
\end{figure*}

Graphical representations of the magnetic order of a single magnetic
moment of $\Euion$ ion (orange solid arrow), its component parallel
to the \cax{} (light blue dashed arrow), external magnetic field
(green dashed-dotted arrow) and total magnetic component parallel
to the \cax{} (dark blue solid arrow) in comparison to the magnetic
phase diagrams of $\Euxii$ (a and b) and $\Euviii$ (c and d) are
presented in Fig.~\ref{fig:Diag}.

When comparing $\Euxii$ and $\Euviii$ (or other Eu-122 systems with
$0.07<x(\ce{Co})<0.2$) one has to have in mind that the microscopic
magnetic structure on Eu-sublattice is different for these two compounds.
As shown by the \MS{} and neutron scattering, the angle between the
magnetic moment of Eu and the \cax{} $\vartheta$ changes with Co-doping
\citep{Blachowski2011Codoped,Jin2016}. Although for both $\Euxii$
and $\Euviii$ their ground state is canted antiferromagnetic (C-AF),
based on \MS{} study, it is known that  in zero-magnetic field the
angle $\vartheta$ is smaller for the Ca-and-Co-doped compound ($\vartheta\in\left\langle 44^{\circ},60^{\circ}\right\rangle $)
than for the Co-only-doped one with $\vartheta\approx30^{\circ}$
\citep{Komedera2018,Tran2017,Blachowski2011Codoped} \textemdash{}
schematically presented in Fig.~\ref{fig:Diag} as step (1). Therefore,
locally the magnetic component parallel to the \cax{} associated
with the magnetic moment of $\Euion$ is bigger for $\Euxii$ than
for $\Euviii$ (see the blue dashed arrows). When magnetic field is
applied perpendicularly to the \cax{} (Fig.~\ref{fig:Diag}~b and~d),
the magnetic moment reorients, finally reaching the field induced
ferromagnetic order (FI-F) where all moments are perpendicular to
the \cax{} and the component parallel to the \cax{} is not present
\textemdash{} Fig.~\ref{fig:Diag} b and d, steps (3a) and (3b).
While the field induced ferromagnetic state is also reached when the
magnetic field is applied parallel to the \cax, this is a completely
different magnetic order, where the magnetic moments of $\Euion$
are parallel to the \cax{} and there is no perpendicular to the \cax{}
component \textemdash{} Fig.~\ref{fig:Diag} a and c, steps (3a)
and (3b).

Another observation that can be made from Fig.~\ref{fig:Diag} is
that superconductivity (SC) appears only in certain regions, especially
when the total magnetic field parallel to the \cax{} is reduced.
Superconductivity exists ``more easily'' when there is no magnetic
field parallel to the \cax{} \textemdash{} either the external applied
field or the field associated with the magnetic moment of $\Euion$
\textemdash{} so when there is no magnetic field perpendicular to
the \ce{FeAs}-layers.

\bibliography{Supplement_1-1}